\documentclass[12pt]{article}
\usepackage{setspace}
\usepackage{amsmath}
\usepackage{amssymb}
\usepackage{abceqn}
\usepackage{graphicx}
\usepackage{times}
\usepackage[margin=1.6cm]{geometry}
\usepackage[round,numbers,sort&compress]{natbib} 
\title{General coarse-grained red blood cell models: I. Mechanics}

\author{Dmitry~Fedosov \\
        Division of Applied Mathematics, \\
        Brown University, Providence, RI
        \and Bruce~Caswell \\
        Division of Engineering, \\
        Brown University, Providence, RI
        \and George~Em~Karniadakis\thanks{
           Corresponding author.  Address: 
           Division of Applied Mathematics,
           Brown University, 
           182 George Street,
           Providence, RI 02912, U.S.A.,
	   Tel.: (401)863-1217, Fax: (401)863-3369} \\
	Division of Applied Mathematics, \\
        Brown University, Providence, RI}

\date{14.04.09}

\begin{document}

\begin{titlepage}
\maketitle
\thispagestyle{empty}
\end{titlepage}

\abstract{We present a rigorous procedure to derive coarse-grained red blood cell (RBC) models, which lead to accurate 
mechanical properties of realistic RBCs. Based on a semi-analytic theory linear and non-linear elastic properties 
of the RBC membrane can be matched with those obtained in optical tweezers stretching experiments.   
In addition, we develop a nearly stress-free model which avoids a number of pitfalls of existing RBC models, 
such as non-biconcave equilibrium shape and dependence of RBC mechanical properties on the triangulation 
quality. The proposed RBC model is suitable for use in many existing numerical methods, such as Lattice Boltzmann, 
Multiparticle Collision Dynamics, Immersed Boundary, etc. 

}

\section{Introduction}
\label{sec:introduction}

Recent experiments on the red blood cell (RBC) to probe its mechanical properties include micropipette aspiration 
\cite{Evans_NMC_1973,Discher_MMR_1994} and RBC deformation by optical tweezers
\cite{Henon_DSM_1999,Mills_NEV_2004,Suresh_CBS_2005}. These experiments provide clear evidence 
that RBCs subject to deformations are characterized by a non-linear mechanical response.
The healthy human RBC assumes a biconcave shape with an average diameter of $7.8$ $\mu m$. RBCs 
have relatively simple structure \cite{Byers_VPA_1985,Liu_VHL_1987} comprising of a membrane filled 
with a liquid cytosol of fixed volume. The RBC membrane consists of a lipid bilayer with an attached 
cytoskeleton formed by a spectrin protein network linked by short actin filaments. The lipid bilayer can be 
considered nearly viscous and area-incompressible \cite{Fung_MPT_1993}, while the attached spectrin network
is mainly responsible for the membrane elastic response providing RBC integrity as it undergoes severe 
deformations in narrow capillaries as small as $3$ $\mu m$ in diameter. This relatively simple RBC 
structure can be considered as an excellent system to study and rigorously model its complex behavior and response.  
To this end, a number of numerical models have been developed recently, including continuum descriptions 
\cite{Fung_MPT_1993,Evans_MTB_1980,Pozrikidis_NSC_2005,Eggleton_LDR_1998} and discrete approximations at 
the spectrin molecular level \cite{Discher_SEC_1998,Li_SLM_2005} as well as at the mesoscopic scale 
\cite{Noguchi_STV_2005,Dupin_MDP_2007,Dzwinel_DPM_2003,Pivkin_ACG_2008}. Fully continuum (fluid and solid) 
type of modeling often suffers from difficulties in coupling nonlinear solid motions and fluid flow and 
from excessive computational expense. Therefore, ``semi-continuum'' modeling \cite{Pozrikidis_NSC_2005,Eggleton_LDR_1998} 
of deformable particles is developing rapidly and is typically based on the immersed boundary or front-tracking 
techniques. Here a membrane is represented by a set of points which move in a Lagrangian fashion and 
are coupled to an Eulerian discretization of fluid domain. Continuum models often do not consider  
membrane thermal fluctuations present at the mesoscopic and microscopic scales. On the other hand, detailed 
spectrin molecular modeling of RBCs is limited due to extreme computational demands. Therefore, in this work 
we will focus on accurate mesoscopic modeling of RBCs. 

There exist several mesoscopic methods \cite{Noguchi_STV_2005,Dupin_MDP_2007,Dzwinel_DPM_2003,Pivkin_ACG_2008} 
for modeling deformable particles and in particular RBCs. Dzwinel et al. 
\cite{Dzwinel_DPM_2003} model RBC as a volume of elastic material which has an inner skeleton. This model 
does not take into account the main structure concept of RBC (a membrane filled with a fluid), 
and therefore it cannot capture the proper dynamics, for example the tumbling and tank-treading behavior 
in shear flow \cite{Abkarian_SSF_2007,Skotheim_RBC_2007}. The other three aforementioned methods 
\cite{Noguchi_STV_2005,Dupin_MDP_2007,Pivkin_ACG_2008} employ a very
similar approach where the RBC is represented
by a network of springs in combination with bending rigidity and constraints for surface-area and volume 
conservation. Dupin et al. \cite{Dupin_MDP_2007} couple the discrete
RBC to a fluid described by the Lattice 
Boltzmann method \cite{Succi_LBM_2001}. Their results are very promising, however the main disadvantage 
of their model is the fact that thermal fluctuations are not considered, which are of importance in RBC rheology and 
dynamics. Noguchi and Gompper \cite{Noguchi_STV_2005} employed Multiparticle Collision Dynamics \cite{Malevanets_MSM_1999} 
and presented encouraging results on vesicles and RBCs. Pivkin and Karniadakis \cite{Pivkin_ACG_2008} used Dissipative 
Particle Dynamics (DPD) \cite{Hoogerbrugge_SMH_1992} for a coarse-grained RBC model which we 
will use as the starting point of our work. Specifically, we will develop a generalized RBC model with major improvements in 
its mechanical properties.                                           
               
The paper is organized as follows. In the next section we present details of the RBC model. 
Section \ref{sec:mechanics} provides a semi-analytical theory of the RBC membrane elastic properties 
and compares the RBC stretching deformation with experimental data. We conclude in section \ref{sec:summary} 
with a brief discussion and detailed suggestions for model development.

\section{Red blood cell model}
\label{sec:rbc}

The membrane model structure is analogous to the works in \cite{Noguchi_STV_2005,Dupin_MDP_2007,Pivkin_ACG_2008}. It is defined 
as a set of points with Cartesian coordinates $\{{\bf x}_i\}$, $i \in 1...N_v$, that are vertices in a two-dimensional 
triangulated network on the RBC surface. The vertices are connected by $N_s$ edges represented by springs, 
which form $N_t$ triangles. The free energy of the system is given by 
\begin{equation}
V(\{ {\bf x}_i \}) = V_{in-plane} + V_{bending} + V_{area} + V_{volume}.
\end{equation}

The in-plane free energy term includes the energy of springs, $U_s$, and may contain the elastic energy stored in the membrane as follows
\begin{equation}
V_{in-plane} = \sum_{j \in 1...N_s} U_s(l_j) + \sum_{k \in 1...N_t} \frac{C_q}{A_k^q},
\label{eq:inplane}
\end{equation}
where $l_j$ is the length of the spring $j$, $A_k$ is the area of the k-th triangle, and the constant $C_q$ and exponent $q$ should 
be properly selected. Different spring models can be used here and we will discuss performance of some of them in section \ref{sec:mechanics}. 
However, we would like to highlight two {\em nonlinear} spring models: the wormlike chain (WLC) and the finitely extensible nonlinear 
elastic (FENE) spring, the attractive potentials of which are given, respectively, by              
\begin{equation}
U_{WLC} = \frac{k_BT l_{max}}{4p} \frac{3x^2-2x^3}{1-x},~~~~~~~~~~~
U_{FENE} = -\frac{k_s}{2}l^2_{max} \log{\left[ 1-x^2 \right]},
\end{equation}
where $x=l/l_{max} \in (0,1)$, $l_{max}$ is the maximum spring extension, $p$ is the persistence length and  $k_s$ is the FENE spring 
constant. Note that when the distance between two connected points approaches $l_{max}$, the corresponding spring force goes to infinity, 
and therefore limits the maximum extension to $l_{max}$. It is important to point out that both WLC and FENE springs exert purely 
attractive forces, thus they produce a triangle area compression, while the second term in equation (\ref{eq:inplane}) provides  
a triangle area expansion. The stressless (or minimum energy) state corresponds to an equilibrium spring length $l_0$ and depends 
on the introduced parameters. The relation among these parameters and the equilibrium length can be derived by an energy minimization 
argument \cite{Discher_SEC_1998} or by setting the Cauchy stress obtained from the virial theorem to zero \cite{Dao_MBA_2006}. 
We obtained the following expressions for WLC and FENE springs, respectively,
\begin{equation}     
C_q^{WLC} = \frac{\sqrt{3} A_0^{q+1} k_BT (4x_0^2-9x_0+6)}{4pql_{max}(1-x_0)^2},~~~~~~~~~
C_q^{FENE} = \frac{\sqrt{3} A_0^{q+1} k_s}{q(1-x_0^2)}, 
\end{equation}        
where $x_0=l_0/l_{max}$ and $A_0 = \sqrt{3} l_0^2/4$. These formulas allow us to calculate the strength of the second term in equation 
(\ref{eq:inplane}) for the given equilibrium length and spring parameters. Another choice is to consider a spring with a specific 
equilibrium length (e.g., harmonic spring, WLC or FENE in combination with a repulsive potential), and then set $C_q$ to zero.
We now introduce a repulsive force defined as a power function (POW) of the separation distance $l$ as follows 
\begin{equation} 
f_{POW}(l) = \frac{k_p}{l^m},~~~~~~~ m>0,
\label{eq:pow} 
\end{equation}        
where $k_p$ is the force coefficient and $m$ is the exponent. The combination of WLC or FENE with POW defines a spring with an 
equillibrium length, and will be called WLC-POW and FENE-POW, respectively. The strength $k_p$ can be expressed in terms of 
the equillibrium length $l_0$ and the WLC or FENE parameters by equating the corresponding forces. The combination of WLC or FENE with
the in-plane energy in equation (\ref{eq:inplane}) will be denoted as WLC-C and FENE-C throughout the paper.

The bending energy is defined as 
\begin{equation}
V_{bending} = \sum_{j \in 1...N_s} k_b \left[ 1-cos(\theta_j-\theta_0) \right],
\end{equation}
where $k_b$ is the bending constant, $\theta_j$ is the instantaneous angle between two adjacent triangles having the common edge $j$,
and $\theta_0$ is the spontaneous angle. 

The area and volume conservation constraints are
\abceqnbeg
\begin{equation}
V_{area} = \frac{k_a (A-A_0^{tot})^2}{2A_0^{tot}} + \sum_{j \in 1...N_t} \frac{k_d (A_j-A_{0})^2}{2A_{0}},
\end{equation}
\begin{equation}
V_{volume} = \frac{k_v (V-V_0^{tot})^2}{2V_0^{tot}},
\end{equation}
\abceqnend
where $k_a$, $k_d$ and $k_v$ are the global area, local area and volume constraint constants, respectively. The terms $A$ and $V$  are the 
total area and volume of RBC, while $A_0^{tot}$ and $V_0^{tot}$ are the {\em desired} total area and volume, respectively. 
Note, that the above expressions define the global area and volume constraints while the second term in equation (7a) 
corresponds to local area dilatation. 

In order to obtain the forces corresponding to the above energies we use the following formula    
\begin{equation}
{\bf f}_i = -\partial V(\{ {\bf x}_i \})/ \partial {\bf x}_i,~~~~~~~~~ i \in 1...N_v. 
\end{equation}
Exact force expressions can be derived analytically from the defined energies, however for brevity we do not present them in this paper.

\section{Mechanical properties}
\label{sec:mechanics}

Mechanical properties of RBCs were measured in a number of experiments by micropipette aspiration
\cite{Evans_NMC_1973,Discher_MMR_1994} and RBC deformation by optical tweezers
\cite{Henon_DSM_1999,Mills_NEV_2004,Suresh_CBS_2005}. The reported shear modulus $\mu_0$ lies between $2$ and $15$ 
$\mu N/m$ and the bending modulus $k$ is between $1\times 10^{-19}$ and $7\times 10^{-19}$ $J$, which corresponds 
to the range of $23-163$ $k_BT$ based on the normal body temperature $T=36.6^o C$. In addition to some uncertainties
in the experiments, the discrepancies in the measurements arise in part from applying simplified geometrical models to 
extract values from the measured forces as the precise geometry is often not known. In such cases, accurate numerical 
modeling can provide a valuable aid in experimental parameter quantification.              

In recent optical tweezers stretching experiments \cite{Mills_NEV_2004,Suresh_CBS_2005} the RBC behavior was 
modeled using a hyperelastic material model and the finite element method (FEM). From the FEM simulations 
the range for the membrane shear modulus of $\mu_0=5-12$ $\mu N/m$ was obtained. This corresponds to the Young's 
modulus of $Y=3\mu_0=15-36$ $\mu N/m$ due to the three-dimensional membrane model. Dao et al. \cite{Dao_MBA_2006} performed 
coarse-grained molecular dynamics (CGMD) simulations of the spectrin-level cytoskeleton which yielded a worse 
approximation to the experimental stretching response in comparison with FEM. They derived the first-order approximation 
of the shear modulus $\mu_0$ and the area-compression modulus $K$ for a regular hexagonal network of springs expressed through 
spring parameters. Eventhough the shear modulus in the FEM and CGMD simulations was matched, it is clear from figure 
8 of \cite{Dao_MBA_2006} that FEM and CGMD systems have different Young's modulus as the slopes in the linear elastic 
deformation regime are different. In addition, their estimated area-compressibility modulus was $K=2\mu$ which yields 
the Poisson's ratio of $\nu=1/3$, while the membrane was nearly incompressible. However, for an incompressible material 
one can find that $\nu=1$ for a two-dimensional membrane model and $K \to \infty$. We have confirmed that their analytical 
results are correct but they appear to be incomplete because not all model contributions are considered for the 
membrane elastic properties estimation, which can explain the inconsistency found. We will explain this further in the 
next section. 

\subsection{Linear elastic properties}

\begin{figure}[!h]
\centering
\includegraphics*[scale=0.3]{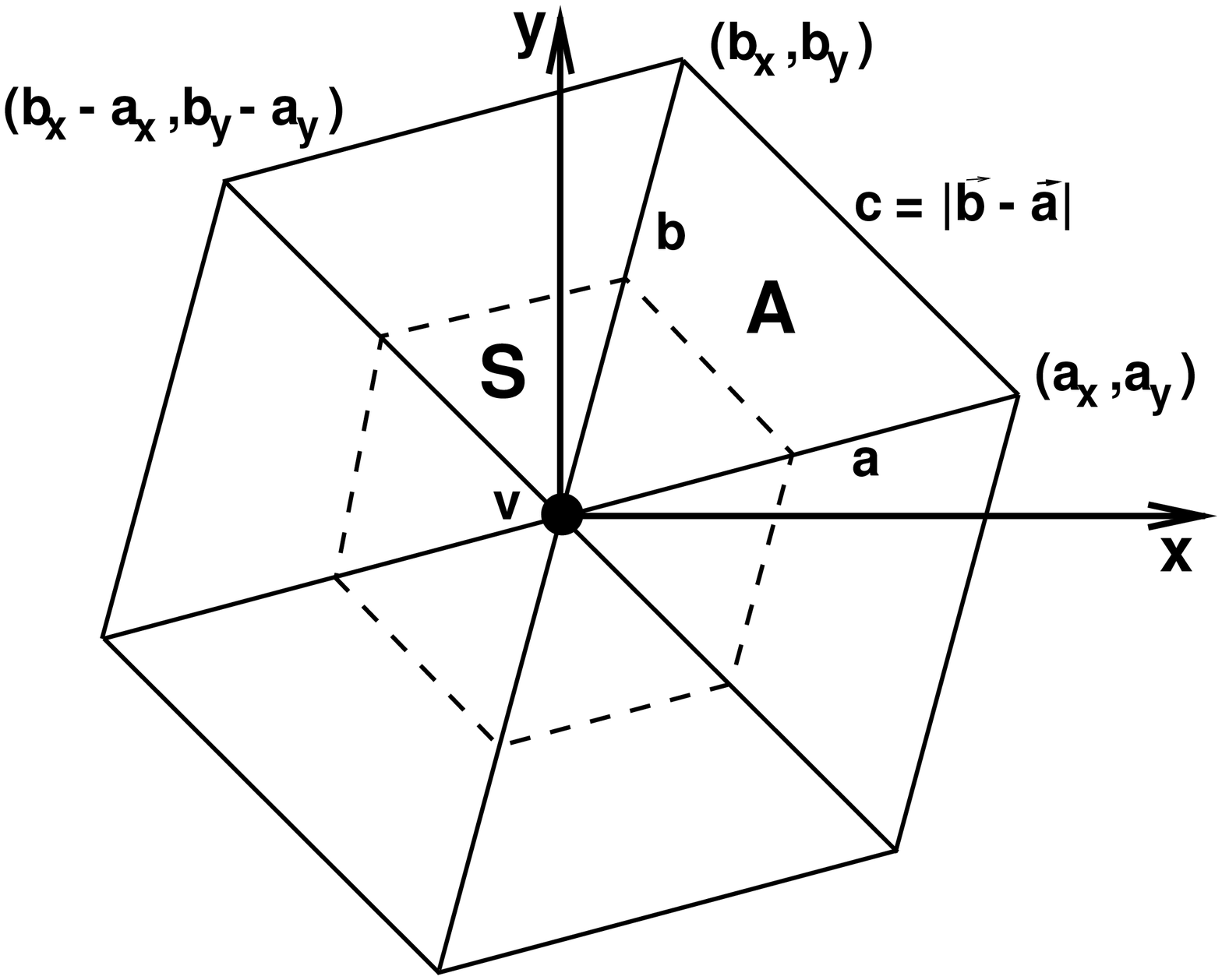}
\includegraphics*[scale=0.4]{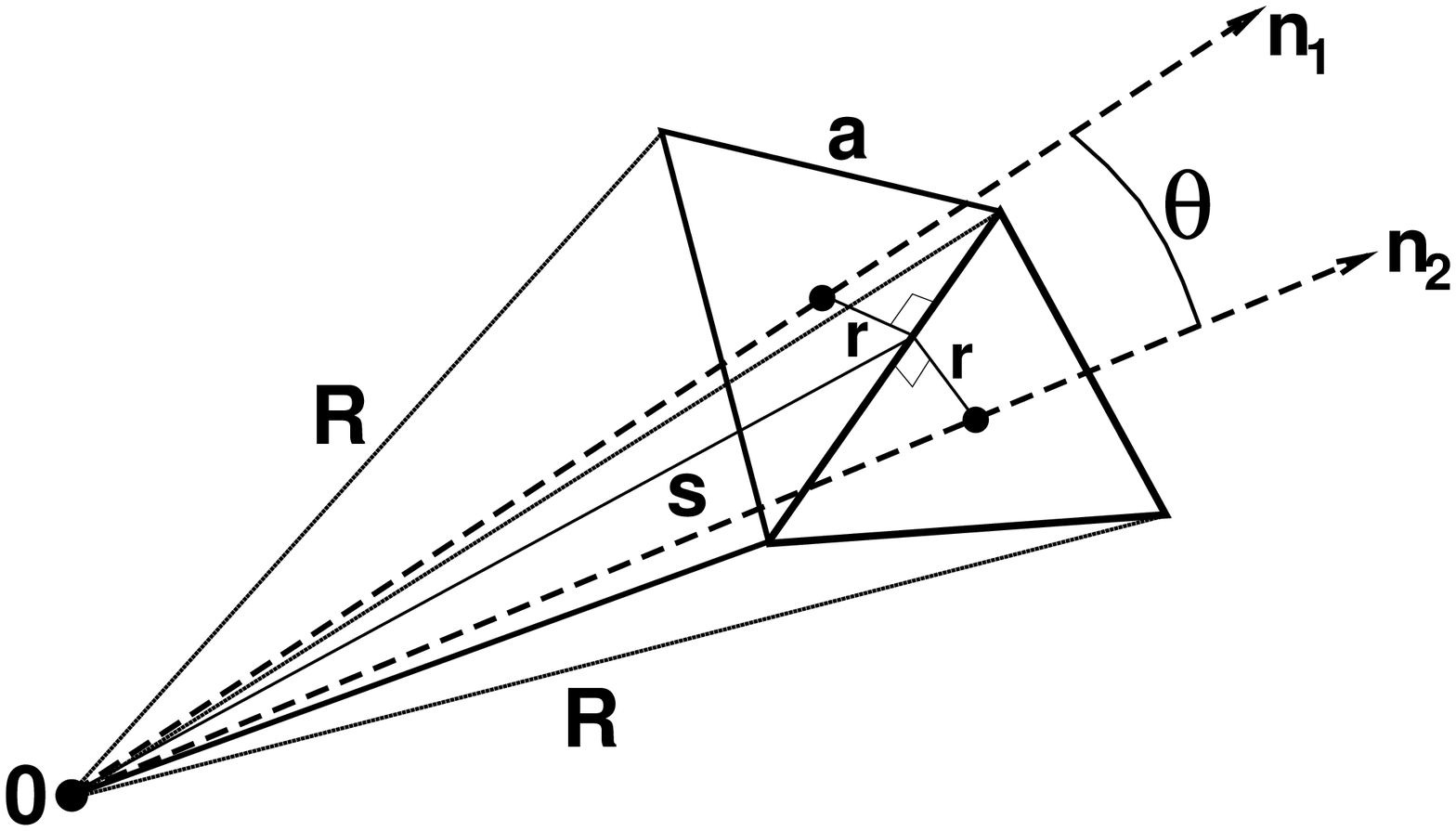}
\caption{}
\label{fig:sketch}
\end{figure}

Our starting point is the linear analysis of a two-dimensional sheet of springs built with equilateral triangles as 
described in detail in \cite{Dao_MBA_2006}. Figure \ref{fig:sketch} (left) shows an element of the equilateral triangulation
with vertex {\bf v} placed at the origin. 
%
%
The stress for the area element $S$ (from the virial theorem) is given by 
\begin{equation*}
\tau_{\alpha \beta} = -\frac{1}{2A} \left [ \frac{f(a)}{a}a_{\alpha}a_{\beta} +  \frac{f(b)}{b}b_{\alpha}b_{\beta} +
\frac{f(c)}{c}(b_{\alpha}-a_{\alpha})(b_{\beta}-a_{\beta}) \right ] -
\end{equation*} 
\begin{equation}
- \left (q \frac{C_q}{A^{q+1}}+\frac{k_a(A_0^{tot}-N_tA)}{A_0^{tot}} + \frac{k_d(A_0-A)}{A_0} \right) \delta_{\alpha \beta},
\end{equation} 
where $f(\cdot)$ is the spring force, $\alpha$, $\beta$ can be $x$ or $y$, $N_t$ is the total number of triangles and 
$A_0^{tot}=N_tA_0$. In general, $N_t$ cancels out and the global and local area contributions to the stress can be combined 
together as $-(k_a+k_d)(A_0-A)/A_0 \delta_{\alpha \beta}$. {\em Note, that the linear analysis in \cite{Dao_MBA_2006} did not take 
into account the global and local area contributions to the stress which significantly affects the final results}.  The linear shear modulus 
can be derived by applying a small engineering shear strain $\gamma$ to the configuration in figure \ref{fig:sketch} (left) and taking the 
first derivative of shear stress $\mu_0=\frac{\partial \tau_{xy}}{\partial \gamma} |_{\gamma=0}$. The shear deformation 
is area-preserving, and therefore only spring forces contribute to the membrane shear modulus. For different spring 
models, we obtained the following expressions for $\mu_0$:               
\abceqnbeg
\begin{equation}
\mu_0^{WLC-C} = \frac{\sqrt{3} k_BT}{4p l_{max} x_0} \left(\frac{3}{4(1-x_0)^2} - \frac{3}{4} + 4x_0 + \frac{x_0}{2(1-x_0)^3} \right),
\label{eq:mu_wlc_c}
\end{equation} 
\begin{equation}
\mu_0^{FENE-C} = \frac{\sqrt{3}k_s}{2} \left(\frac{x_0^2}{(1-x_0^2)^2} + \frac{2}{1-x_0^2} \right),
\label{eq:mu_fene_c}
\end{equation}
\begin{equation}
\mu_0^{WLC-POW} = \frac{\sqrt{3} k_BT}{4p l_{max} x_0} \left( \frac{x_0}{2(1-x_0)^3} - \frac{1}{4(1-x_0)^2} + \frac{1}{4} \right) + 
\frac{\sqrt{3} k_p (m+1)}{4l_0^{m+1}},
\label{eq:mu_wlc_pow}
\end{equation}
\begin{equation}
\mu_0^{FENE-POW} = \frac{\sqrt{3}}{4} \left(\frac{2 k_s x_0^2}{(1-x_0^2)^2} + \frac{k_p (m+1)}{l_0^{m+1}} \right).
\label{eq:mu_fene_pow}
\end{equation}
\abceqnend

The linear elastic area compression modulus $K$ can be calculated from the area expansion with the resulting pressure given by
\begin{equation}
P = -\frac{1}{2}(\tau_{xx}+\tau_{yy})=\frac{3 l f(l)}{4A} + q \frac{C_q}{A^{q+1}}+ \frac{(k_a+k_d)(A_0-A)}{A_0}.
\label{eq:pressure}
\end{equation}
The compression modulus $K$ is defined as 
\begin{equation}
K = -\left. \frac{\partial P}{\partial \log{(A)}} \right|_{A=A_0} = -\left.\frac{1}{2}\frac{\partial P}{\partial \log{(l)}} \right|_{l=l_0}=
-\left.\frac{1}{2}\frac{\partial P}{\partial \log{(x)}} \right|_{x=x_0}.
\label{eq:compression}
\end{equation}
Using equations (\ref{eq:pressure}) and (\ref{eq:compression}) we derive the linear area compression modulus for different spring 
models as follows
\abceqnbeg
\begin{equation}
K^{WLC-C} = \frac{\sqrt{3} k_BT}{4p l_{max}(1-x_0)^2} \left[ \left( q + \frac{1}{2} \right ) (4x_0^2-9x_0+6) +
\frac{1+2(1-x_0)^3}{1-x_0} \right] + k_a + k_d,
\end{equation} 
\begin{equation}
K^{FENE-C} = \frac{\sqrt{3}k_s}{1-x_0^2} \left[ q+1+ \frac{x_0^2}{1-x_0^2} \right] + k_a + k_d,
\end{equation}
\begin{equation}
K^{WLC-POW} = 2\mu_0^{WLC-POW} + k_a + k_d,
\end{equation}
\begin{equation}
K^{FENE-POW} = 2\mu_0^{FENE-POW} + k_a + k_d.
\end{equation}
\abceqnend  
Note, that if $q=1$ we obtain the expressions $K^{WLC-C} = 2\mu_0^{WLC-C} + k_a + k_d$ and $K^{FENE-C} = 2\mu_0^{FENE-C} + k_a + k_d$.
Generally, for a nearly incompressible sheet of springs the area constraint coefficients have to be large such that $k_a+k_d \gg 1$,
and thus $K \gg \mu_0$.

The Young's modulus $Y$ for the two-dimensional sheet can be expressed through the shear and area compression modulus as follows 
\begin{equation}
Y=\frac{4K\mu_0}{K+\mu_0},~~~~~~~ Y  \to 4\mu_0,~~if~~ K \to \infty,   
\label{eq:young_modulus}
\end{equation}
and the Poisson's ratio $\nu$ is given by
\begin{equation}
\nu=\frac{K-\mu_0}{K+\mu_0},~~~~~~~ \nu  \to 1,~~if~~ K \to \infty.   
\end{equation}
The above expressions are consistent with the incompressibility assumption enforced through the condition $k_a+k_d \gg 1$. 
In practice, we use the value of $k_a+k_d=5000$ which provides a nearly incompressible membrane with the Young's modulus 
of about $2\%$ smaller than the asymptotic value of $4\mu_0$ ($\mu_0=100$). All the analytical expressions for $\mu_0$, $K$ and $Y$ were numerically 
verified by shearing, area expanding and stretching experiments of the regular two-dimensional sheet of springs. In addition,
it is important to note that the modeled sheet appears to be {\em isotropic} for small shear and stretch deformations, however 
it is {\em anisotropic} at large deformations.

\subsection{Membrane bending properties}
 
In this section we discuss the correspondence of our bending model to the macroscopic model of Helfrich \cite{Helfrich_EPB_1973} 
given by
\begin{equation}
E = \frac{k_c}{2} \int_A (C_1+C_2-2C_0)^2 dA + k_g \int_A C_1 C_2 dA,
\label{eq:bend}   
\end{equation}
where $C_1$ and $C_2$ are the local principal curvatures, $C_0$ is the spontaneous curvature, and $k_c$ and $k_g$ are the bending rigidities. 

We base the derivation on the spherical shell. Figure \ref{fig:sketch} (right) shows two equilateral triangles with sides $a$, the vertices
of which rest on the surface of a sphere of radius $R$.   
The angle between their normals ${\bf n}_1$ and ${\bf n}_2$ is equal to $\theta$. For the spherical shell we can derive from equation 
(\ref{eq:bend}) $E=8\pi k_c (1-C_0/C_1)^2 + 4\pi k_g= 8\pi k_c
(1-R/R_0)^2 + 4\pi k_g$, where $C_1=C_2=1/R$ and $C_0 = 1/R_0$. For the triangulated sphere 
we have $E_t=N_s k_b [1-cos(\theta-\theta_0)]$ in the defined notations. We expand $cos(\theta-\theta_0)$ in Taylor series around 
$(\theta-\theta_0)$ to obtain $E_t=N_s k_b (\theta-\theta_0)^2/2 + O((\theta-\theta_0)^4)$, where we neglect high-order terms. From 
figure \ref{fig:sketch} (right) we find that $2r \approx \theta R$ or $\theta = \frac{a}{\sqrt{3}R}$, and analogously 
$\theta_0 = \frac{a}{\sqrt{3}R_0}$. Furthermore,  $A_{sphere}=4\pi R^2 \approx N_tA_0=\frac{\sqrt{3}N_t a^2}{4}=
\frac{N_s a^2}{2\sqrt{3}}$, and thus $a^2/R^2=8\pi \sqrt{3}/N_s$. Finally, we obtain $E_t=N_s k_b (\frac{a}{\sqrt{3}R} - 
\frac{a}{\sqrt{3}R_0})^2/2= \frac{N_s k_b a^2}{6 R^2}(1-R/R_0)^2=\frac{8\pi k_b}{2\sqrt{3}}(1-R/R_0)^2$. Equating the macroscopic bending 
energy $E$ for $k_g=-4k_c/3$, $C_0=0$ \cite{Lidmar_VSB_2003} and $E_t$ gives us the relation $k_b=2k_c/ \sqrt{3}$, which is the same as 
derived in the continuum limit in \cite{Lidmar_VSB_2003}.
The spontaneous angle $\theta_0$ is set according to the total number of vertices $N_v$ on the sphere. It can be shown that 
$\cos{(\theta)}=1-\frac{1}{6(R^2/a^2-1/4)}=(\sqrt{3}N_s-10\pi)/(\sqrt{3}N_s-6\pi)$, while $N_s=2N_v-4$. The corresponding bending stiffness 
$k_b$ and the spontaneous angle $\theta_0$ are given by 
\begin{equation}
k_b=\frac{2}{\sqrt{3}}k_c,~~~~~~~~ \theta_0=cos^{-1} \left( \frac{\sqrt{3}(N_v-2)-5\pi}{\sqrt{3}(N_v-2)-3\pi} \right).
\label{eq:bend_angle}
\end{equation}

\subsection{RBC triangulation}

The average unstressed shape of a single red blood cell measured in the experiments in \cite{Evans_MTB_1980} is biconcave and is described by
\begin{equation}
z=\pm D_0 \sqrt{1-\frac{4(x^2+y^2)}{D_0^2}}\left[ a_0 + a_1\frac{x^2+y^2}{D_0^2} + a_2\frac{(x^2+y^2)^2}{D_0^4} \right],
\label{eq:rbc}
\end{equation}
where $D_0=7.82$ $\mu m$ is the cell diameter, $a_0=0.05179025$, $a_1=2.002558$, and $a_2=-4.491048$. The area and volume of this 
RBC is equal to $135$ $\mu m^2$ and $94$ $\mu m^3$, respectively. We have investigated three types of triangulation 
strategies: 
\begin{itemize}
  \item {\em Point charges}: $N_v$ points are randomly distributed on the sphere surface, and the electrostatic problem
of point charges is solved while the point movements are constrained on the sphere. After steady state is reached the sphere 
surface is triangulated and conformed to the RBC shape according to equation (\ref{eq:rbc}). 
  \item {\em Gridgen}: the RBC shape is imported into commercially available grid generation software Gridgen \cite{gridgen} which performs the advancing front method for the RBC surface triangulation. 
  \item {\em Energy relaxation}: First, the RBC shape is triangulated following the point charges or Gridgen methods. 
Subsequently, the relaxation of the free energy of the RBC model is performed while the vertices are restricted to move on the biconcave shape in equation (\ref{eq:rbc}). The relaxation procedure includes only in-plane and bending energy components and is done by flipping between the two diagonals of two adjacent 
triangles. 
\end{itemize}
The triangulation quality can be characterized by two distributions: (i) distribution of the link (edge) length, 
(ii) distribution of the vertex degrees (number of links in the vertex junction). The former is characterized by the value 
$d(l)=\sigma(l)/\bar{l}$, where $\bar{l}$ is the average length of all edges, and $\sigma(l)$ is the standard deviation. The latter 
defines the regularity of triangulation by providing the relative percentage of degree-$n$ vertices $n=1...n_{max}$. Note that 
the regular network, for which the mechanical properties were derived, has only  degree-6 vertices. Table \ref{tab:mesh_quality}
presents the mesh quality data {\em on average} for different triangulation methods. 
%
%
The better mesh quality corresponds to a combination of smaller $d(l)$, higher percentage of degree-6, and smaller percentage of 
any other degree vertices, and is achieved for larger number of points $N_v$. At this point, it seems that the best quality is 
reached with the free energy relaxation method while the worst is the Gridgen (advancing front method) triangulation, which will be discussed further below.

\subsection{Dissipative Particle Dynamics modeling and scaling to real units}

We will model RBC with Dissipative Particle Dynamics (DPD), a mesoscale method used for simulations of complex fluids and soft matter, see \cite{Hoogerbrugge_SMH_1992} for details. We now outline the scaling procedure which relates DPD non-dimensional units to real units. First, we choose the 
equilibrium spring length $l_0 = l_0^D$ in DPD units, and the superscript $D$ denotes ``DPD'' and $[l_0^D]=r_c$, where
$r_c$ defines DPD length scale. Another 
parameter we are free to select is the imposed shear modulus $\mu_0=\mu_0^D$ with $[\mu_0^D]=\frac{N^D}{r_c}=
\frac{(k_BT)^D}{r_c^2}$, which will provide a scaling base. Use of WLC and FENE springs requires to set the maximum extension
length $l_{max}^D$, however it is more convenient to set the ratio $r_{mult}=l_0^D/l_{max}^D$. Further in the paper we will show that 
the choice of $r_{mult}$ does not affect the linear elastic deformation, but it governs the RBC non-linear response at large deformation. 
For given $l_0^D$, $\mu_0^D$  and $r_{mult}$ we can calculate the required spring 
parameters for a chosen model using equations (10a-d). Then, the area-compression modulus $K^D$ and the Young's modulus $Y^D$
are found for the calculated spring parameters and given area constraint parameters ($k_a$ and $k_d$) using equations 
(13a-d, \ref{eq:young_modulus}). At this point, we can define the length scaling based on the cell diameter $D_0^D=(L_x^D+L_y^D)/2$,
where $[D_0^D]=r_c$ and $L_x$, $L_y$ are the cell diameters in $x$ and $y$ directions found from the equilibrium simulation of 
a single cell using the previously obtained model parameters. The length scaling based on $l_0^D$ appears to be inappropriate, 
because, in general, the cell dimensions would depend on the relative volume to area ratio and to some extent on the present 
triangulation artifacts (discussed later in text). As an example, we can define RBC and a spherical vesicle with the same 
$l_0^D$, while the cell sizes would greatly differ. However, in general, $D_0^D$ is proportional to $l_0^D$ for fixed volume 
to area ratio. The real RBC has the average diameter $D_0^R=7.82$ $\mu m$ (superscript $R$ denotes ``real''), and 
therefore the following length scaling is adapted
\begin{equation}
r_c =\frac{D_0^R}{D_0^D} [m].
\end{equation}

Due to the fact that we will perform RBC stretching simulations, it is natural to involve the Young's modulus into the 
scaling as the main parameter. Matching the real and model Young's modulus $Y^D$ $\frac{(k_BT)^D}{r_c^2}$ = 
$Y^R$ $\frac{(k_BT)^R}{m^2}$ provides us with the energy unit scaling as follows 
\begin{equation}
(k_BT)^D = \frac{Y^R}{Y^D} \frac{r_c^2}{m^2} (k_BT)^R = \frac{Y^R}{Y^D} \left( \frac{D_0^R}{D_0^D} \right)^2 (k_BT)^R.
\label{eq:temp_scale}
\end{equation}
After we determined the DPD energy unit (as an example for the human body temperature of $T=36.6^o C$), we can calculate 
the bending rigidity in DPD using the energy unit and equation (\ref{eq:bend_angle}). In addition, we define the force scaling by 
\begin{equation}
N^D = \frac{(k_BT)^D}{r_c} = \frac{Y^R}{Y^D} \frac{D_0^R}{D_0^D} \frac{(k_BT)^R}{m}=N^R.
\label{eq:force_scale}
\end{equation}
Note that for the stretching simulations here we do not need to explicitly define mass and time scaling as we are not 
interested in stretching dynamics.

\subsection{RBC stretching: success and problems}

We perform RBC stretching simulations and compare results with the experimental data of RBC deformation by 
optical tweezers \cite{Suresh_CBS_2005}. Here, we assume that the real RBC has diameter $D_0^R=7.82$ $\mu m$. The aforementioned 
FEM simulations of RBC membrane  \cite{Suresh_CBS_2005} showed an agreement with the experimental data 
for $\mu_0^R=5.3$ $\mu N/m$, however we find that a slightly better correspondence of the results is achieved 
for $\mu_0^R=6.3$ $\mu N/m$ and $Y^R=18.9$ $\mu N/m$, which we select to be the targeted properties. Table \ref{tab:wlc_c_params} 
shows a set of the required RBC parameters.     
%
%
The triangulation for all $N_v$ was performed using the free energy relaxation method. The imposed Young's modulus for 
all cases is $Y^D = 392.453$, which is about $2\%$ lower than that in the incompressible limit $Y^D=4\mu_0^D=400$. 
Using equation (\ref{eq:temp_scale}) we find the energy unit $(k_BT)^D$ based on $(k_BT)^R$ at the normal body 
temperature $T=36.6^o C$. The bending rigidity $k_c$ is set to $2.4 \times 10^{-19} J$, which seems to be a widely accepted 
value and is equal to approximately $56(k_BT)^R$. The total RBC area $A_0^{tot}$ is equal to $N_t\frac{\sqrt{3}}{4}(l_0^D)^2$, 
where $N_t$ is the total number of triangle plaquettes with the area $A_0=\frac{\sqrt{3}}{4}(l_0^D)^2$. Note that for all 
triangulations used in this paper $N_t=2N_v-4$. The total RBC volume $V_0^{tot}$ is found according to the following scaling 
$V_0^{tot}/(A_0^{tot})^{3/2} = V^R/(A^R)^{3/2}$, where $V^R=94$ $\mu m^3$ and  $A^R=135$ $\mu m^2$ according to the average 
RBC shape described by equation (\ref{eq:rbc}).   

The modeled RBC is suspended in a solvent which consists of free DPD particles with number density $n=3$. Note that macroscopic 
solvent properties (e.g., viscosity) are not important here, because we are interested in the final cell deformation for every 
constant stretching force. Thus, we allow enough time for the RBC to reach its final deformation state without close monitoring of 
the stretching dynamics. Meanwhile, the solvent maintains the
temperature at the constant value of $(k_BT)^D$.  

\begin{figure}[!h]
\centering
\includegraphics*[scale=0.27]{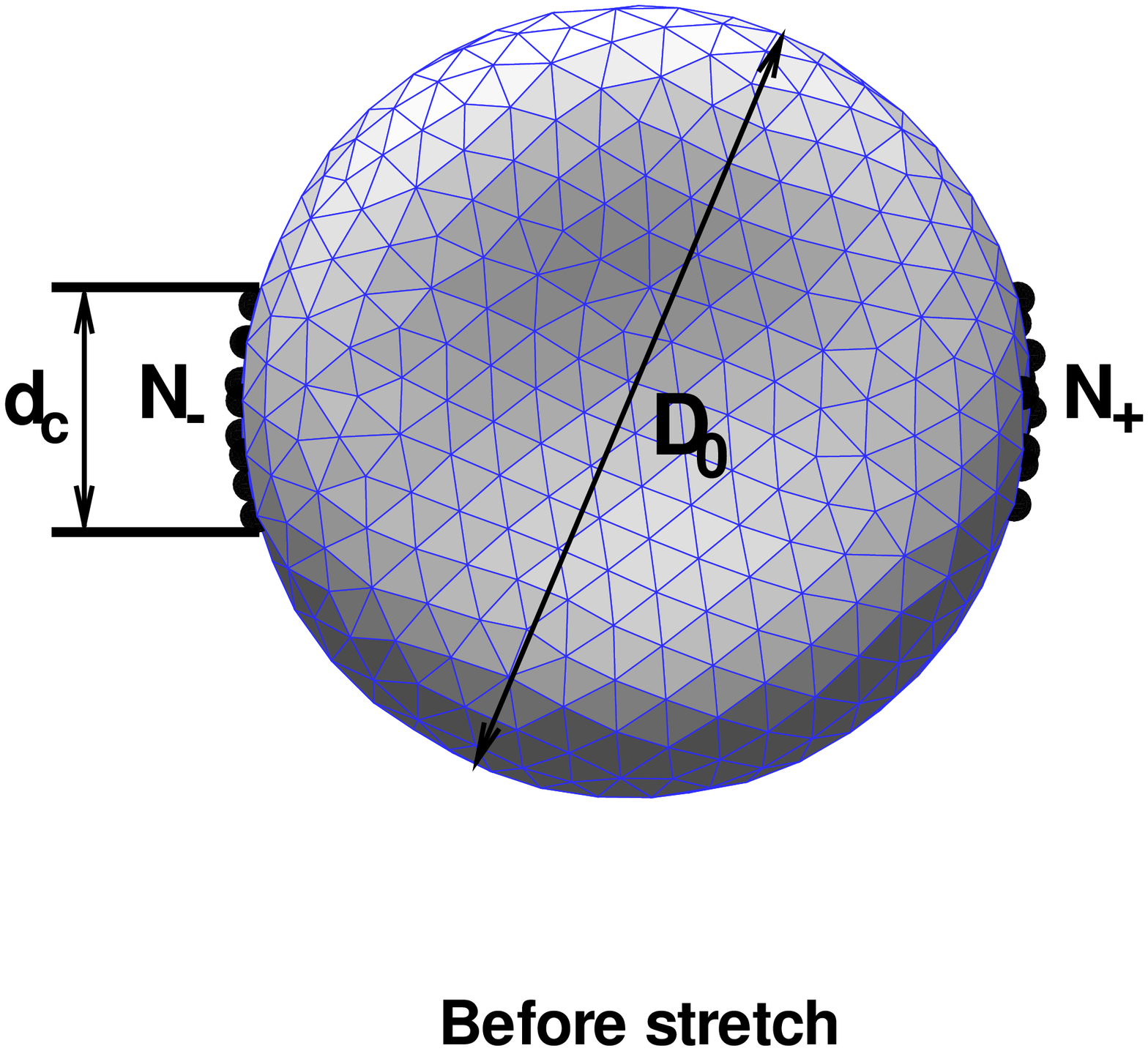}
\includegraphics*[scale=0.27]{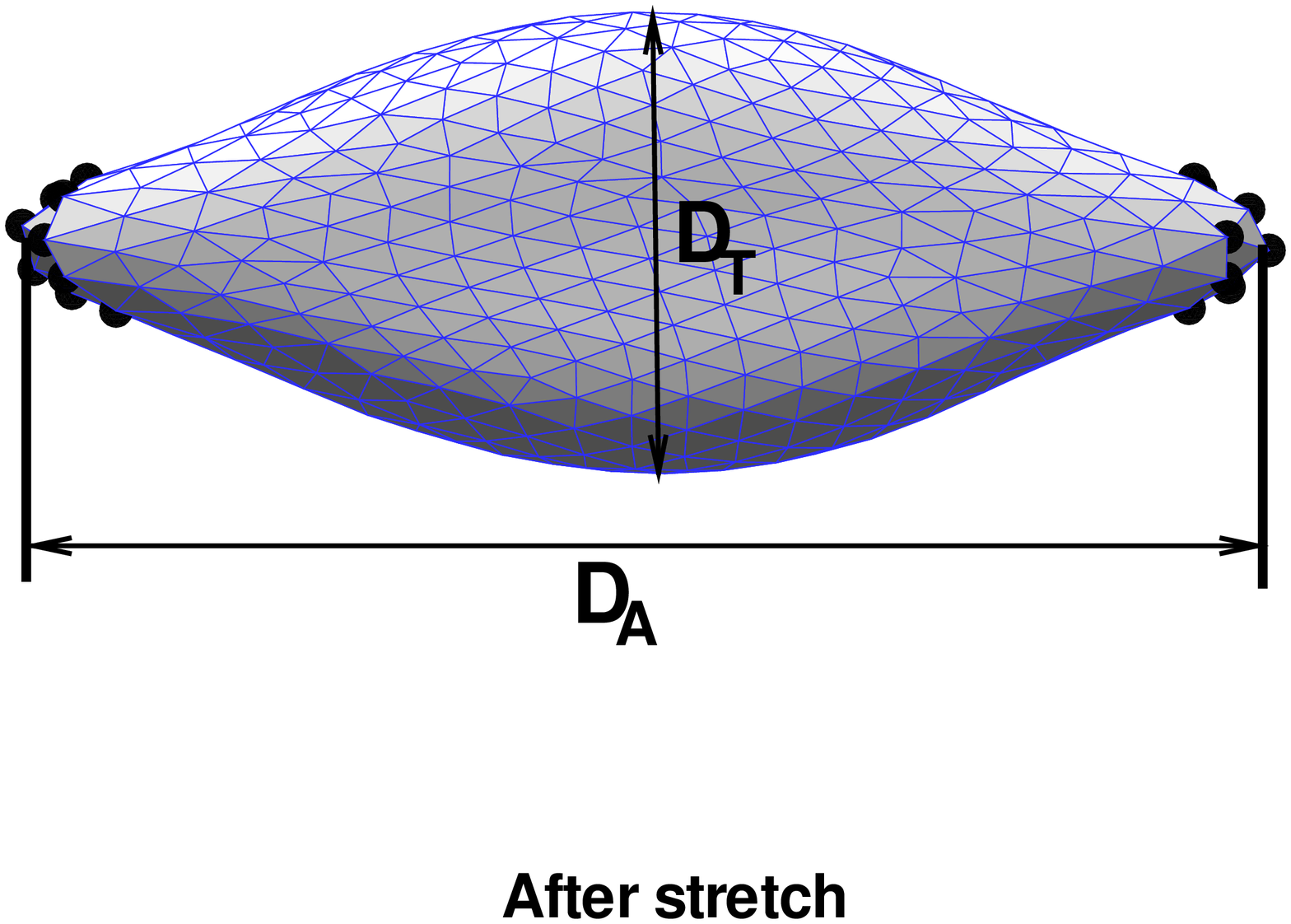}
\caption{}
\label{fig:rbc_sketch}
\end{figure}

Figure \ref{fig:rbc_sketch} shows a sketch of the red blood cell before and after deformation.  
%
%
The total stretching force $F_s^R$ is in the range $0...200$ $pN$, which can be scaled into DPD units $F_s^D$ according to 
equation (\ref{eq:force_scale}).
The total force $F_s^D$ is applied to $N_+=\epsilon N_v$ vertices (drawn as small black spheres in figure \ref{fig:rbc_sketch}) 
of the membrane with the largest x-coordinates in the positive x-direction, and correspondingly $-F_s^D$ is exerted on 
$N_-=N_+$ vertices with the smallest x-coordinates in the negative x-direction. Therefore, a vertex in $N_+$ or $N_-$ is 
subject to the force $f_s^D=\pm F_s^D/(\epsilon N_v)$. The vertex fraction $\epsilon$ is equal to $0.02$ corresponding 
to a contact diameter of the attached silica bead $d_c=2$ $\mu m$ used in experiments. The contact diameter was measured as 
$\left( \max_{ij}|y_i^+-y_j^+|+\max_{ij}|y_i^--y_j^-| \right)/2$, where $y_i^+$, $y_j^+$ and $y_i^-$, $y_j^-$ are the 
y-coordinates of vertices in $N_+$ and $N_-$, respectively. The simulations for the given force range were performed 
as follows: (i) $M=16$ is chosen, which defines the force increment $\Delta F_s^R = 200~pN/M$ with corresponding 
$\Delta F_s^D$. (ii) The loop $i=1...M$ is run with the stretching force $i\cdot \Delta F_s^D$ during time $2\tau$ each.
The time $\tau$ is long enough in order for RBC to converge to the final stretching state for the given force. Thus, 
the time $[0,\tau]$ is the transient time for convergence, and during time $[\tau,2\tau]$ the deformation response is 
calculated. The axial diameter $D_A$ is computed over time $\tau$ as $|x_{max}-x_{min}|$, where $x_{max}$ is the maximum $x$ 
position among the $N_+$ vertices, while $x_{min}$ is the minimum among $N_-$. The transverse diameter $D_T$ is calculated 
as $2 \times \max_{i=1...N_v} \sqrt{(y_i-c_y)^2 + (z_i-c_z)^2}$, where $c_y$, $c_z$ are the $y$ and $z$ center of mass
coordinates.               
   
\begin{figure}[!h]
\centering
\includegraphics*[scale=0.3]{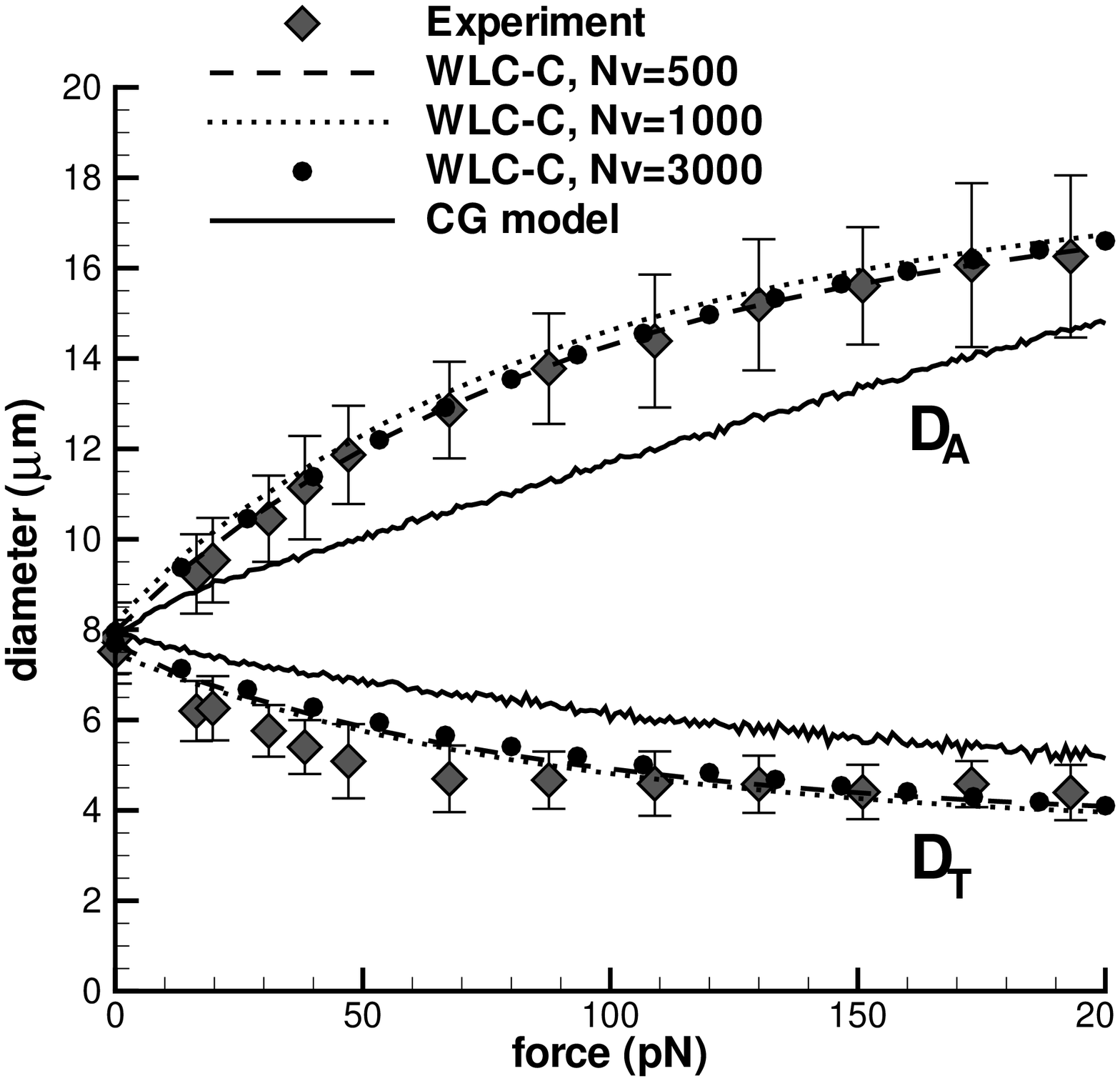}
\includegraphics*[scale=0.3]{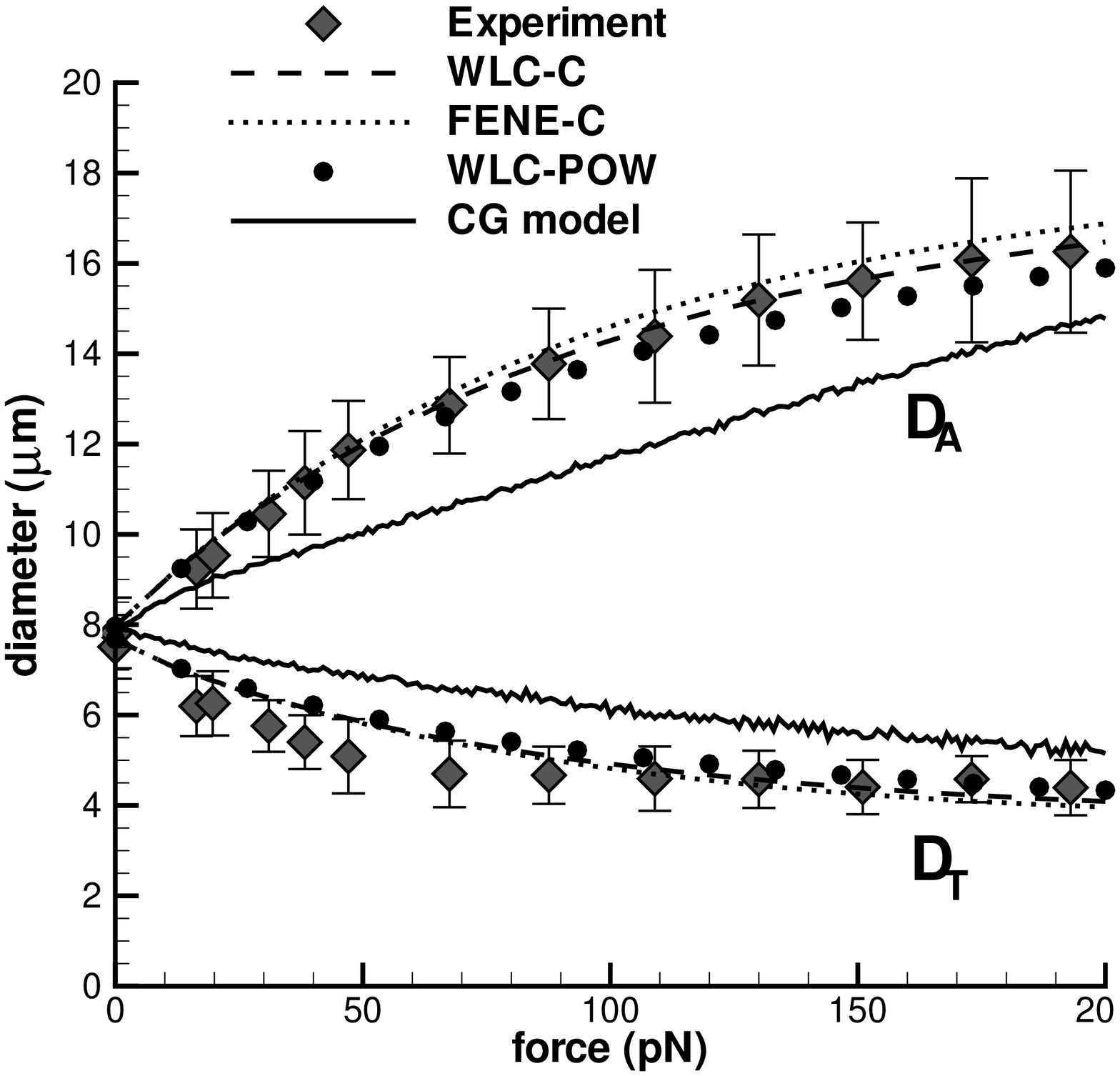}
\caption{}
\label{fig:rbc_stretch}
\end{figure}

Figure \ref{fig:rbc_stretch} presents the RBC stretching response for different number of vertices $N_v$ (left) and spring models 
(right) with RBC parameters from table \ref{tab:wlc_c_params}; also included are experimental results \cite{Suresh_CBS_2005} and
the coarse-grained RBC model results of \cite{Pivkin_ACG_2008}. 
%
%
We find an excellent agreement of the simulation results with the experiment independently of the number of vertices or 
spring model. A noticeable disagreement in the transverse diameter may be partially due to the experimental errors 
rising from the fact that the optical shape measurements were performed from a single observation angle. RBCs subjected 
to stretching may rotate in y-z plane which was noticed in numerical simulations, and therefore measurements done from 
a single observation angle may result in underprediction of the maximum transverse diameter. However, the simulation 
results remain within the experimental error bars. The solid line in figure \ref{fig:rbc_stretch} corresponds to 
the coarse-grained RBC \cite{Pivkin_ACG_2008} of similar type employing the WLC-C model. In \cite{Pivkin_ACG_2008} the derivation of 
linear elastic properties did not include a contribution of the area constraint, which results in Young's modulus 
underprediction on the order of $50\%$. From the region of small near-linear deformation ($0-50$ $pN$) it is clear 
that the solid line corresponds to a membrane with a larger Young's modulus compared to the experiment. In addition, 
in order to compensate for the error in the estimated membrane elastic properties the ratio $r_{mult}$ was set to $3.17$,
which results in near-linear elastic deformation, and ignores a non-linear RBC response at large deformations.       
Finally, it is worth commenting that the FENE-C model appears to be less stable (requires a smaller time step) at 
large deformation due to a more rapid spring hardening compared to WLC-C. The WLC-POW model performs similar to 
WLC-C, however a weak local area conservation ($k_d>0$) may be required for stability at large deformations as it 
mimics the second in-plane force term in equation (\ref{eq:inplane}) for the WLC-C model.  

\begin{figure}[!h]
\centering
\includegraphics*[scale=0.3]{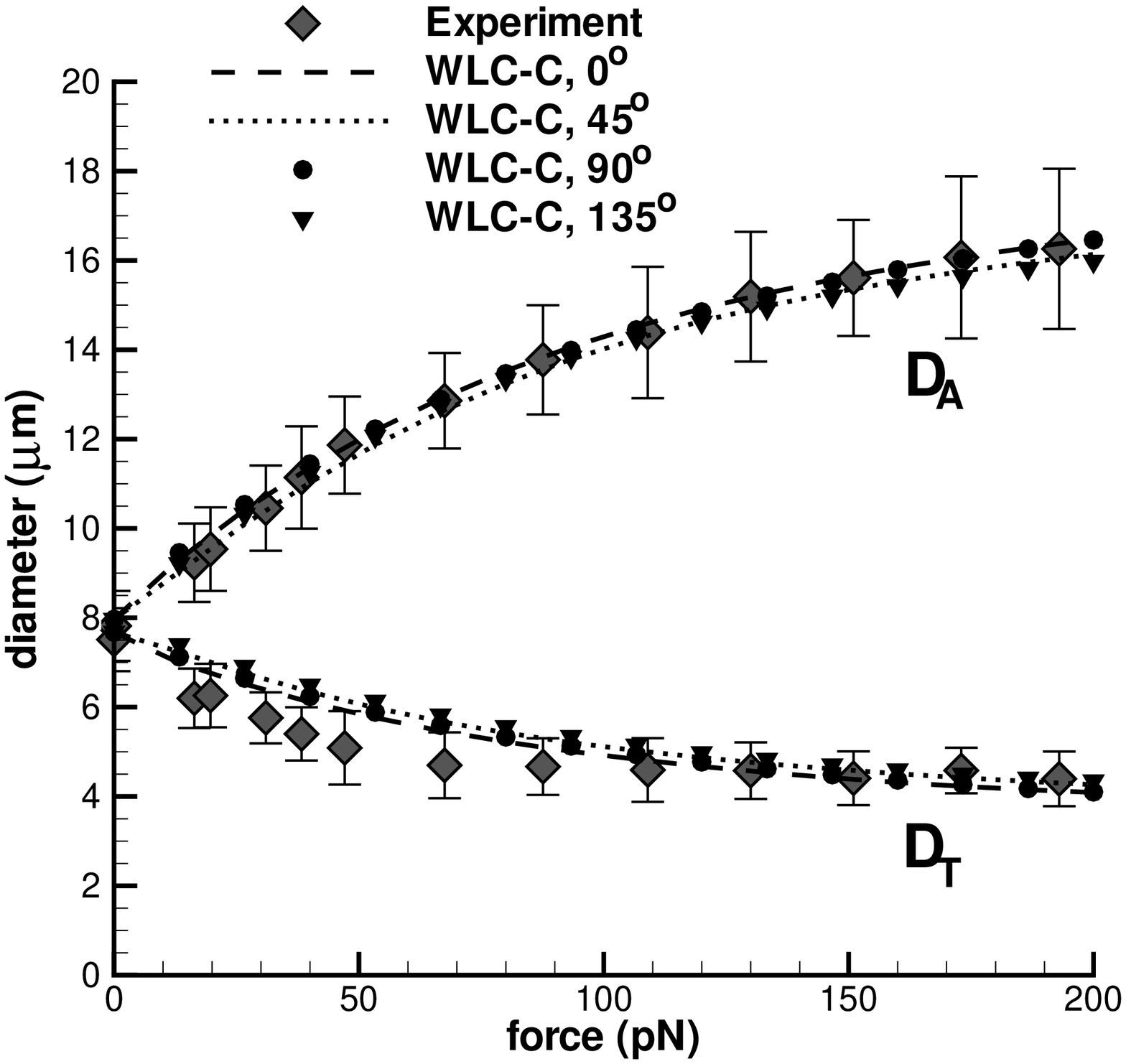}
\includegraphics*[scale=0.3]{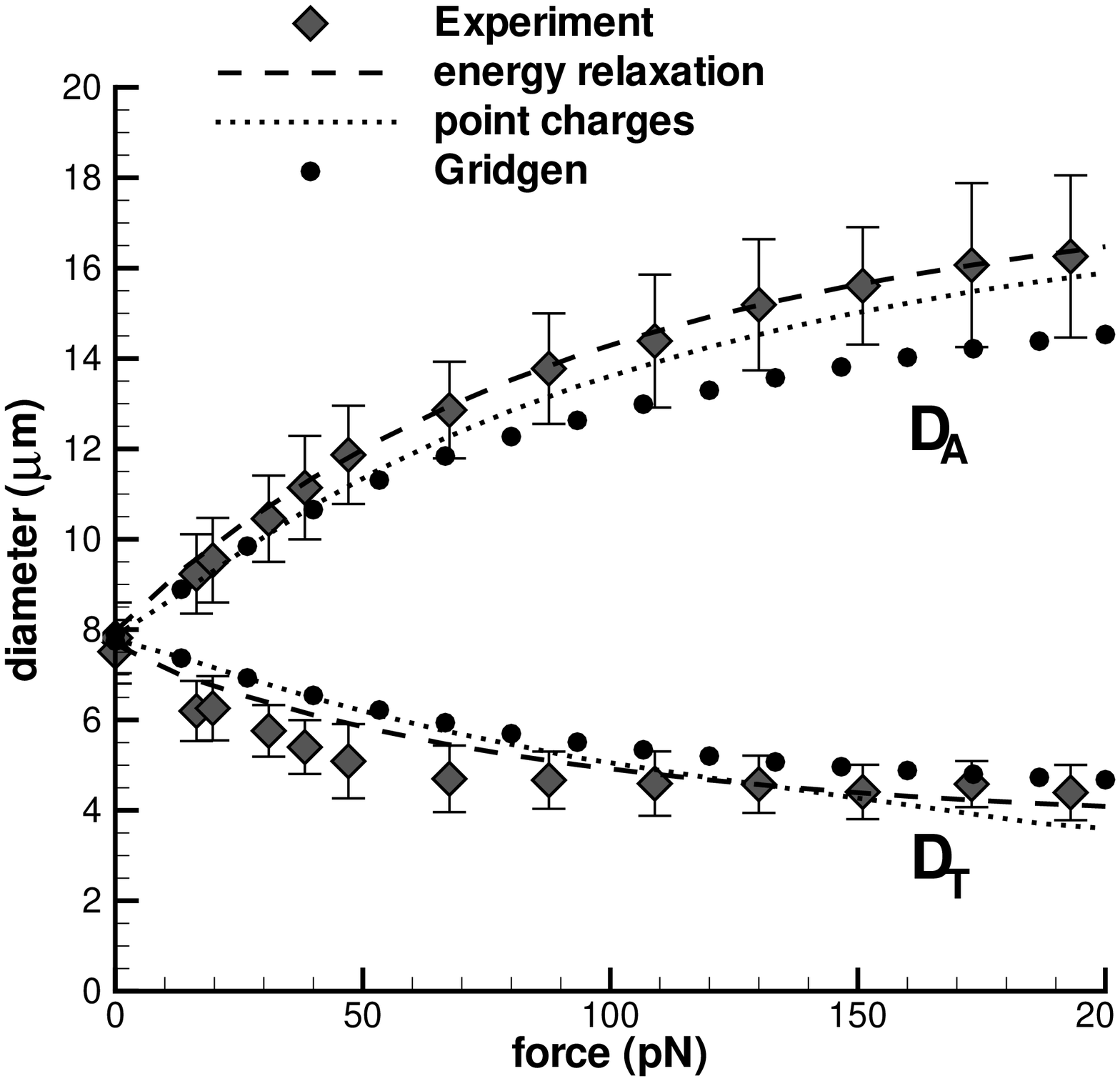}
\caption{}
\label{fig:rbc_problems}
\end{figure}

Despite the demonstrated success of the RBC models, several problems are remaining due to a not stress-free membrane. Figure 
\ref{fig:rbc_problems} shows the RBC response for different triangulation methods: free energy relaxation 
triangulated WLC-C $N_v=500$ RBC stretched along lines with distinct orientation angles (left) and RBC response 
for models with different triangulation (right).
%
%
While RBC triangulated through free energy relaxation method gives satisfactory results with difference in the stretching 
response on the order of $5-8\%$, RBCs triangulated by other methods show much greater discrepancy with the experiment. 
In addition to that, RBCs triangulated by point charges and Gridgen methods require to set the bending rigidity to $300(k_BT)^R$
and $200(k_BT)^R$, respectively, in order to maintain the equilibrium biconcave shape, while the bending rigidity of the real RBC 
is about $56(k_BT)^R$. Here, a lower bending rigidity may result in relaxation to the stomatocyte (cup) shape \cite{Li_SLM_2005}.
Moreover, figure \ref{fig:rbc_problems} shows that these models have higher effective elastic modulus than that predicted as 
they are subject to a higher membrane stress at equilibrium due to triangulation artifacts. Also, they appear to give a stronger 
stretching anisotropy ($10-15\%$) compared to the free energy relaxation method.

\subsection{Stress-free membrane model}

To eliminate the aforementioned membrane stress artifacts we propose a simple modification to the described model. For each 
spring we define $l_0^i$ $i=1...N_s$ set to the spring lengths after the shape triangulation, since 
we assume it to be the RBC equilibrium state. We define accordingly $l_{max}^i=l_0^i\times r_{mult}$ and $A_0^j$ 
$j=1...N_t$ for each triangular plaquette. The total RBC area $A_0^{tot}=\sum_{j=1...N_t} A_0^j$ and the 
total volume $V_0^{tot}$ is calculated from the RBC triangulation. Then, we define the average spring length 
$\bar{l}_0=\sum_{i=1...N_s} l_0^i$ and the average maximum spring extension as $\bar{l}_{max}=\bar{l}_0\times r_{mult}$, 
which are used in the linear elastic properties estimation using equations (14c,d and 17c,d). Here, we omit the WLC-C and 
FENE-C models because it may not be possible to define a single in-plane area expanding potential (the second force term 
in equation (\ref{eq:inplane})) for a triangle with distinct sides. However, for the WLC-POW and FENE-POW models 
the individual equilibrium spring length can be simply defined. Thus, the WLC and FENE spring parameters ($k_BT/p$ and $k_s$)
will be the same for all springs calculated through $\bar{l}_0$, $\bar{l}_{max}$ and $\mu_0^D$, while the power force 
coefficient $k_p$ in equation (\ref{eq:pow}) will be adjusted in order to set the given equilibrium spring length.  
               
\begin{figure}[!h]
\centering
\includegraphics*[scale=0.3]{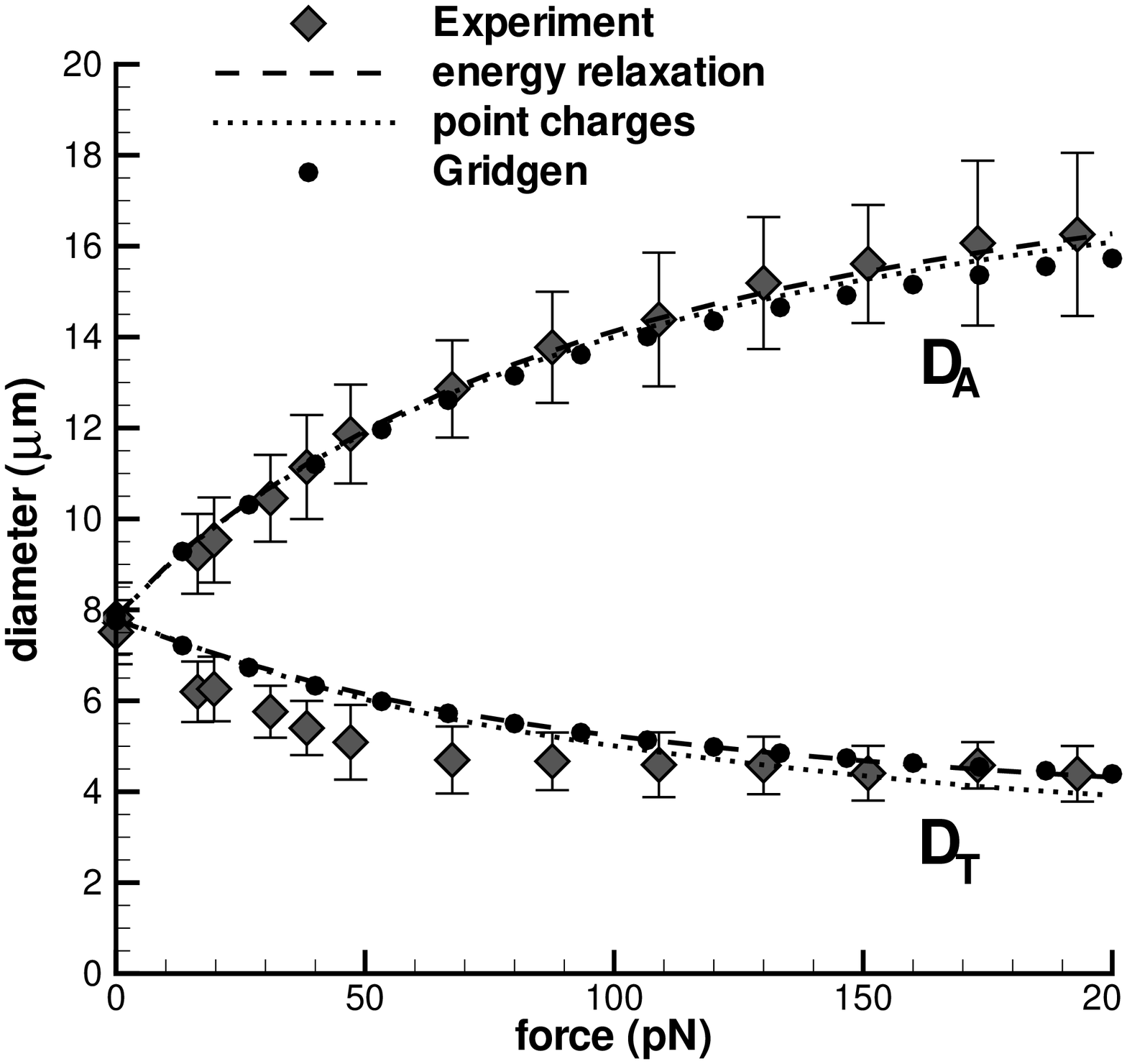}
\includegraphics*[scale=0.3]{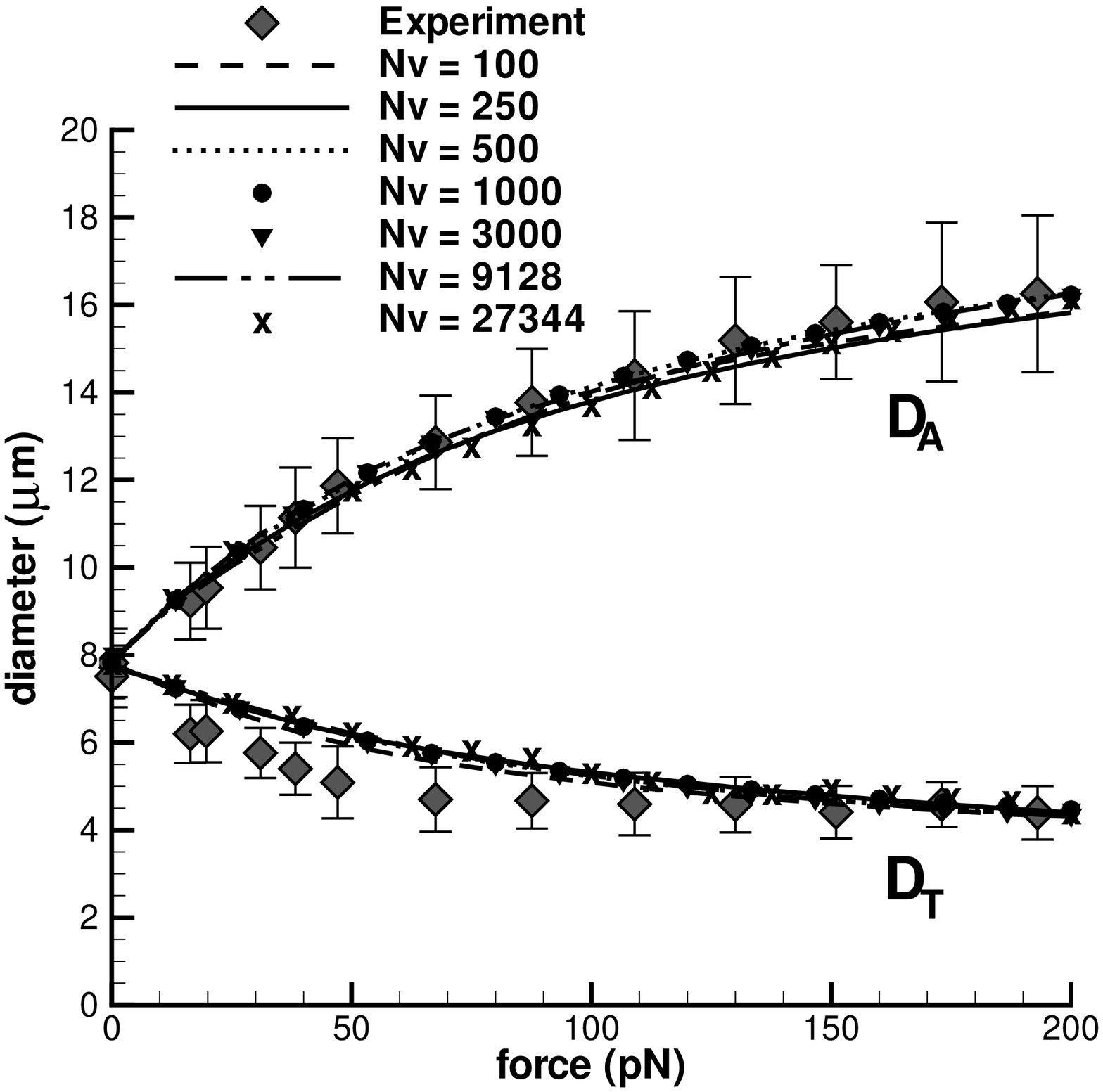}
\caption{}
\label{fig:rbc_new_model}
\end{figure}

We perform tests using the WLC-POW model for different triangulation methods and number of vertices. Figure 
\ref{fig:rbc_new_model} presents simulation results for $N_v=500$ with different triangulations (left) and 
a range of the number of vertices $N_v$ from $100$ to $27344$ (right).
%
%
A substantial improvement is observed when compared with the results in figure \ref{fig:rbc_problems} (right). Note that 
the stress-free model, when probed along different stretching directions results in deviation in the stretching response
on the order of $1\%$ for the free energy triangulation method and about $3-5\%$ for the other triangulation techniques.   
The stretching response for different number of vertices gives an excellent agreement of the results with the experiment.
Here, $N_v=27344$ corresponds to a spectrin-level of RBC modeling as in \cite{Li_SLM_2005}, while $N_v=100-500$ is highly 
coarse-grained RBC. Eventhough the coarse-grained model of $N_v=100$ yields correct mechanical deformation results, 
it may not provide an accurate smooth RBC shape description which can be of importance for the dynamics. We propose that 
the minimum $N_v$ to be used for the RBC model should be about $250-300$. 
   
\begin{figure}[!h]
\centering
\includegraphics*[scale=0.3]{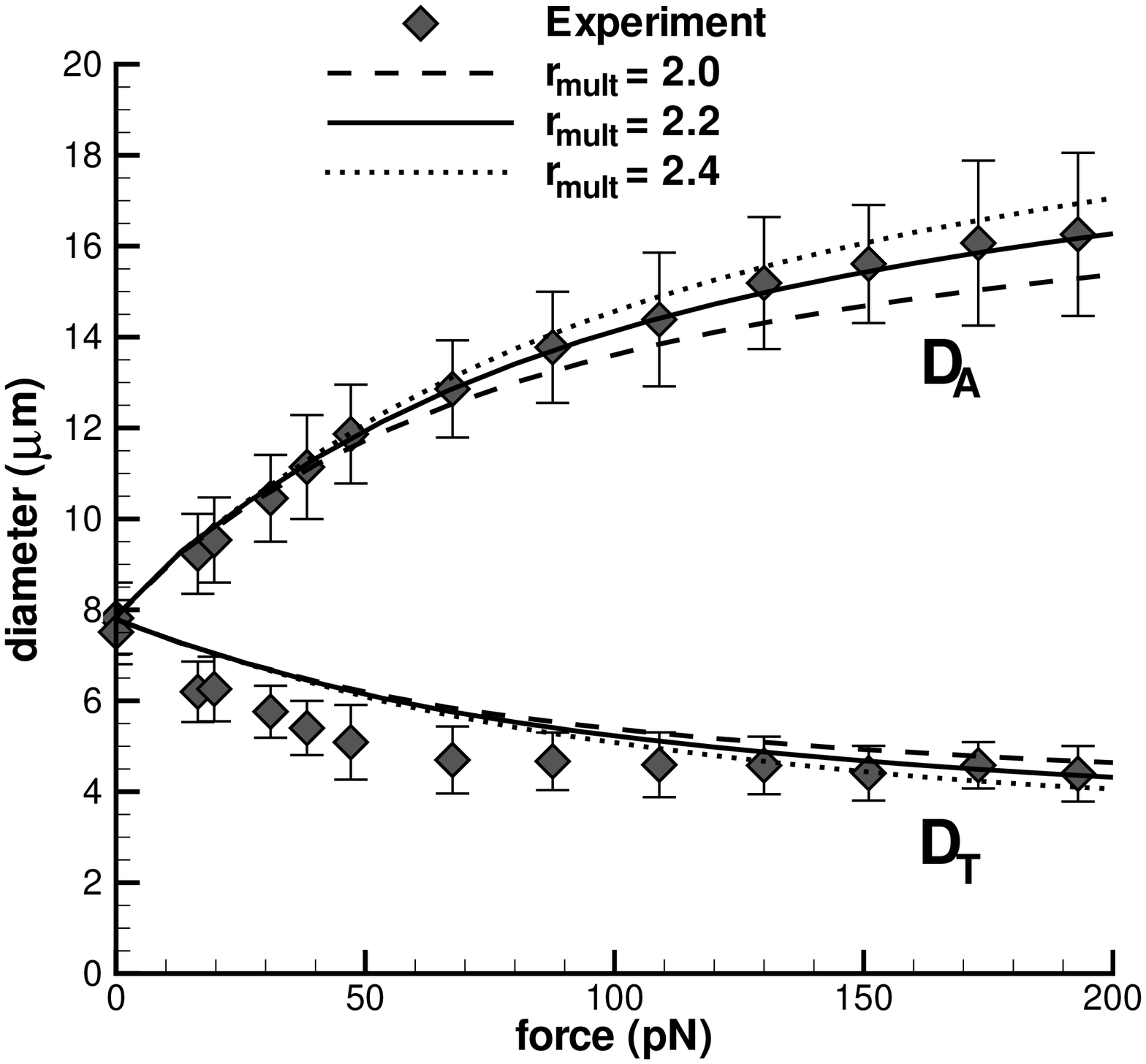}
\includegraphics*[scale=0.3]{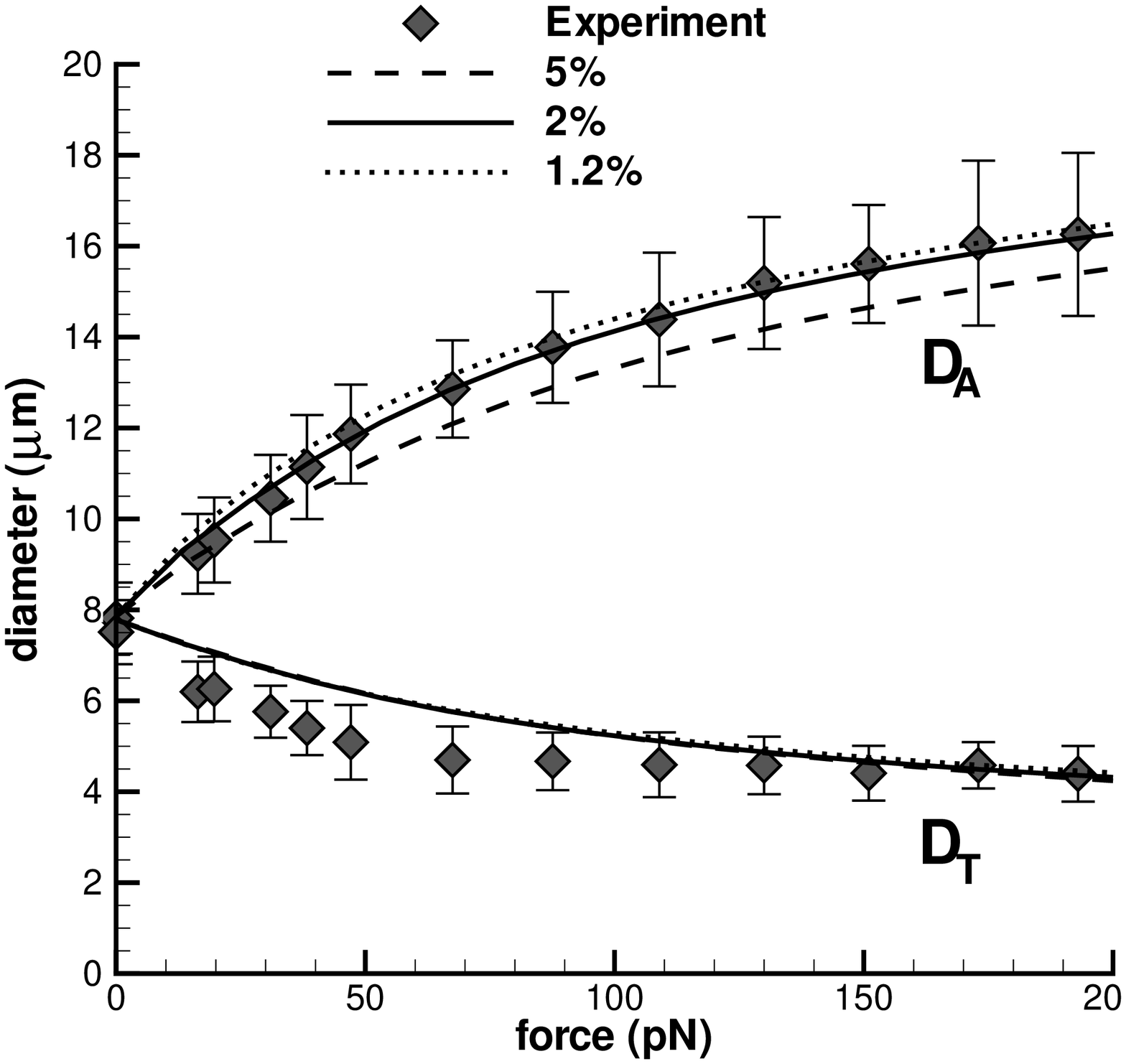}
\caption{}
\label{fig:rbc_dependence}
\end{figure}

Dependence of the RBC deformation response on the ratio $r_{mult}$ and on the number of vertices $N_+$, $N_-$ (figure \ref{fig:rbc_sketch})  
is shown in figure \ref{fig:rbc_dependence}. 
%
%
As we mentioned before small RBC deformations are independent of the ratio $r_{mult}$, however at large deformation this 
parameter plays a significant role and governs the non-linear RBC response. In addition, figure \ref{fig:rbc_dependence}
(right) shows that the RBC response is sensitive to the fraction of vertices (shown in percent) to which the stretching force 
is applied. It is equivalent to changing $d_c$ in figure \ref{fig:rbc_sketch}, which characterizes the attachment area 
of silica bead in the experiments.

\subsection{Comparison with a single spectrin tetramer}

\begin{figure}[!h]
\centering
\includegraphics*[scale=0.4]{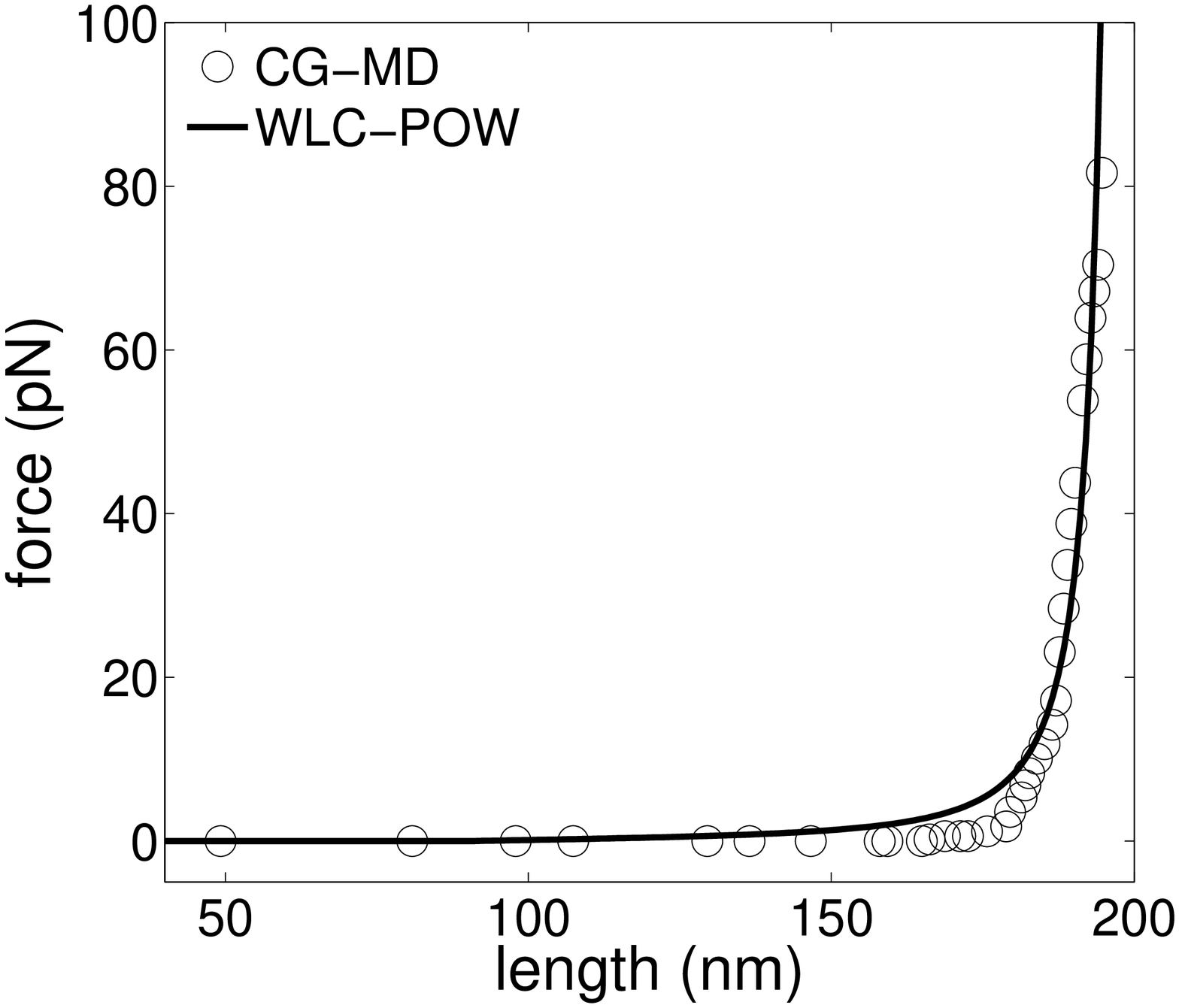}
\caption{}
\label{fig:spectrin}
\end{figure}

It is rather remarkable that RBCs can be accurately modeled with just a few hundred points, which is about hundred 
times computationally cheaper than the spectrin-level RBC model, where $N_v \sim 27000$ \cite{Dao_MBA_2006}. At the spectrin-level of RBC modeling, each 
spring represents a single spectrin tetramer, and therefore the spring force WLC-POW should mimic the spectrin 
tetramer deformation response. We are not aware of any experimental single spectrin stretching results, however this has been done by 
means of numerical simulation using a coarse-grained molecular dynamics (CG-MD) approach in \cite{Voth_UEP_2008}. Figure 
\ref{fig:spectrin} presents a single spectrin tetramer stress-strain response compared with the spring force of 
the spectrin-level RBC model.       
%
%
We find a remarkable agreement of the results. Here, we assume that the maximum extension spring length is $200$ $nm$ as 
in the CG-MD simulations of \cite{Voth_UEP_2008}. This corresponds to $l_0=91$ $nm$ with $r_{mult}=2.2$ from where an 
appropriate length and force scaling can be calculated as done in the previous sections.

\section{Summary}
\label{sec:summary}

We presented general coarse-grained RBC models represented by a network of springs in 
combination with bending rigidity, area, and volume conservation constraints. The modeled RBC accurately 
captures elastic response at small and large deformations, and agrees very well with the 
experiments of the RBC stretching with optical tweezers. The linear elastic properties of the RBC 
membrane are derived analytically, and therefore no manual adjustment of the model parameters through 
numerical tests is required. We also proposed a stress-free RBC model which leads to 
triangulation-independent membrane properties, while the conventional RBC model suffers from non-zero 
local stresses which result in triangulation dependent deformation response and equilibrium shape. 
The model was tested for different levels of coarse-graining starting from the spectrin-level modeling 
(about $N_v=27000$ vertices) and ending with only $N_v=100$ vertices for the full membrane representation. 
However, we suggest that the minimum number of vertices to be used for the RBC membrane should be about 
$N_v=250-300$, while the lower $N_v$ may not accurately represent RBC smooth shape which is of importance 
for the RBC dynamics. In addition, we found an excellent agreement of single spring force in case of 
the spectrin-level model with the spectrin tetramer response obtained from the coarse-grained molecular 
dynamics simulations. The proposed model is general enough, and therefore can be easily applied in many 
numerical methods, such as semi-continuum methods (Immersed Boundary and Advanced Front Tracking), 
mesoscopic methods (Lattice Boltzmann and Brownian Dynamics), and mesoscopic particle methods 
(Dissipative Particle Dynamics and Multiparticle Collision Dynamics). 

Here, we summarize the procedure for the RBC model. First, we obtain triangulation of the equilibrium RBC 
shape defined by equation (\ref{eq:rbc}) for the given number of vertices $N_v$. This triangulation sets 
the required equilibrium lengths for the springs, triangle areas and the total RBC area and volume. 
Second, we choose the modeled membrane shear modulus $\mu_0$, and area and volume constraint 
coefficients (eq. (7a,b)). This defines our RBC model parameters using equations 
(10a-d, 13a-d, 14 and 17) with the average equilibrium spring length, and will set a scaling to the real 
units using equations (20,21). In addition, we need to define the length scaling (eq. (19)) based on the 
RBC diameter. We suggest to obtain the RBC diameter through an equilibrium simulation rather than assuming 
it from the analytical RBC shape (eq. (\ref{eq:rbc})) as they may be slightly different depending on the 
relative contributions of in-plane elasticity and membrane bending rigidity. After these two simple steps,
the linear elastic properties of the model will match those of the real RBC. In addition, we need to mention 
that in the case of strong RBC deformations we may need to adjust the spring maximum extension length 
which governs the non-linear RBC response. However, it is convenient to set the ratio $r_{mult}=l_0/l_{max}=2.2$
for the WLC springs and $r_{mult}=2.05$ for the FENE springs. We emphasize that the described 
procedure does not involve any parameter adjustments through a number of numerical tests. 

The spectrin stretching comparison provides additional justification of using the spring model for accurate RBC deformation 
response. From these results we can draw the conclusion that: {\em an appropriate spring model for RBC should have the maximum 
allowed extension length, in the neighborhood of which the spring force rapidly hardens in order to prevent further 
membrane strain}. In view of this, the harmonic spring used in \cite{Noguchi_STV_2005} gives an adequate response 
at small deformations but it will not capture a non-linear RBC deformations. Furthermore, the neo-Hookean spring used 
in \cite{Dupin_MDP_2007} provides a good RBC stretching response but it may fail at very large deformations. At this point, 
an experimental confirmation of the single spectrin tetramer stress-strain relation would be of great interest.

This paper is the first part of our work on accurate coarse-grained RBC modeling. This simple 
coarse-grained model is inexpensive and it accurately captures the RBC mechanical properties. The second 
part of our work concerns RBC rheological properties and dynamics. 
It will demonstrate the importance of treating the RBC membrane as a viscoelastic material, and will present studies of the complex 
RBC dynamics and rheology.



\bibliography{main}

\begin{thebibliography}{27}
\providecommand{\url}[1]{\texttt{#1}}
\providecommand{\urlprefix}{ }

\bibitem[Evans(1973)]{Evans_NMC_1973}
Evans, E.~A., 1973.
\newblock New membrane concept applied to the analysis of fluid shear- and
  micropipette-deformed red blood cells.
\newblock \emph{Biophys. J.} 13:941--954.

\bibitem[Discher et~al.(1994)Discher, Mohandas, and Evans]{Discher_MMR_1994}
Discher, D.~E., N.~Mohandas, and E.~A. Evans, 1994.
\newblock Molecular maps of red cell deformation: hidden elasticity and in situ
  connectivity.
\newblock \emph{Science} 266:1032--1035.

\bibitem[Henon et~al.(1999)Henon, Lenormand, Richert, and
  Gallet]{Henon_DSM_1999}
Henon, S., G.~Lenormand, A.~Richert, and F.~Gallet, 1999.
\newblock A new determination of the shear modulus of the human erythrocyte
  membrane using optical tweezers.
\newblock \emph{Biophys. J.} 76:1145--1151.

\bibitem[Mills et~al.(2004)Mills, Qie, Dao, Lim, and Suresh]{Mills_NEV_2004}
Mills, J.~P., L.~Qie, M.~Dao, C.~T. Lim, and S.~Suresh, 2004.
\newblock Nonlinear elastic and viscoelastic deformation of the human red blood
  cell with optical tweezers.
\newblock \emph{Mech. Chem. Biosys.} 1:169--180.

\bibitem[Suresh et~al.(2005)Suresh, Spatz, Mills, Micoulet, Dao, Lim, Beil, and
  Seufferlein]{Suresh_CBS_2005}
Suresh, S., J.~Spatz, J.~P. Mills, A.~Micoulet, M.~Dao, C.~T. Lim, M.~Beil, and
  T.~Seufferlein, 2005.
\newblock Connections between single-cell biomechanics and human disease
  states: gastrointestinal cancer and malaria.
\newblock \emph{Acta Biomaterialia} 1:15--30.

\bibitem[Byers and Branton(1985)]{Byers_VPA_1985}
Byers, T.~J., and D.~Branton, 1985.
\newblock Visualization of the protein associations in the erythrocyte-membrane
  skeleton.
\newblock \emph{Proc. Natl. Acad. Sci. USA} 82:6153--6157.

\bibitem[Liu et~al.(1987)Liu, Derick, and Palek]{Liu_VHL_1987}
Liu, S.~C., L.~H. Derick, and J.~Palek, 1987.
\newblock Visualization of the hexagonal lattice in the erythrocyte-membrane
  skeleton.
\newblock \emph{J. Cell Biol.} 104:527--536.

\bibitem[Fung(1993)]{Fung_MPT_1993}
Fung, Y.~C., 1993.
\newblock Biomechanics: {M}echanical properties of living tissues.
\newblock Springer-Verlag, New York, second edition.

\bibitem[Evans and Skalak(1993)]{Evans_MTB_1980}
Evans, E.~A., and R.~Skalak, 1993.
\newblock Mechanics and thermodynamics of biomembranes.
\newblock CRC Press, Inc., Boca Raton, Florida.

\bibitem[Pozrikidis(2005)]{Pozrikidis_NSC_2005}
Pozrikidis, C., 2005.
\newblock Numerical Simulation of Cell Motion in Tube Flow.
\newblock \emph{Ann. Biomed. Engin.} 33:165--178.

\bibitem[Eggleton and Popel(1998)]{Eggleton_LDR_1998}
Eggleton, C.~D., and A.~S. Popel, 1998.
\newblock Large deformation of red blood cell ghosts in a simple shear flow.
\newblock \emph{Phys. Fluids} 10:1834.

\bibitem[Discher et~al.(1998)Discher, Boal, and Boey]{Discher_SEC_1998}
Discher, D.~E., D.~H. Boal, and S.~K. Boey, 1998.
\newblock Simulations of the erythrocyte cytoskeleton at large deformation.
  {II}. {M}icropipette aspiration.
\newblock \emph{Biophys. J.} 75:1584--1597.

\bibitem[Li et~al.(2005)Li, Dao, Lim, and Suresh]{Li_SLM_2005}
Li, J., M.~Dao, C.~T. Lim, and S.~Suresh, 2005.
\newblock Spectrin-level modeling of the cytoskeleton and optical tweezers
  stretching of the erythrocyte.
\newblock \emph{Biophys. J.} 88:3707--3719.

\bibitem[Noguchi and Gompper(2005)]{Noguchi_STV_2005}
Noguchi, H., and G.~Gompper, 2005.
\newblock Shape transitions of fluid vesicles and red blood cells in capillary
  flows.
\newblock \emph{Proc. Natl. Acad. Sci. USA} 102:14159--14164.

\bibitem[Dupin et~al.(2007)Dupin, Halliday, Care, Alboul, and
  Munn]{Dupin_MDP_2007}
Dupin, M.~M., I.~Halliday, C.~M. Care, L.~Alboul, and L.~L. Munn, 2007.
\newblock Modeling the flow of dense suspensions of deformable particles in
  three dimensions.
\newblock \emph{Phys. Rev. E} 75:066707.

\bibitem[Dzwinel et~al.(2003)Dzwinel, Boryczko, and Yuen]{Dzwinel_DPM_2003}
Dzwinel, W., K.~Boryczko, and D.~A. Yuen, 2003.
\newblock A discrete-particle model of blood dynamics in capillary vessels.
\newblock \emph{J. Coll. Interface Science} 258:163--173.

\bibitem[Pivkin and Karniadakis(2008)]{Pivkin_ACG_2008}
Pivkin, I.~V., and G.~E. Karniadakis, 2008.
\newblock Accurate coarse-grained modeling of red blood cells.
\newblock \emph{Phys. Rev. Lett.} 101:118105.

\bibitem[Abkarian et~al.(2007)Abkarian, Faivre, and Viallat]{Abkarian_SSF_2007}
Abkarian, M., M.~Faivre, and A.~Viallat, 2007.
\newblock Swinging of red blood cells under shear flow.
\newblock \emph{Phys. Rev. Lett.} 98.

\bibitem[Skotheim and Secomb(2007)]{Skotheim_RBC_2007}
Skotheim, J.~M., and T.~W. Secomb, 2007.
\newblock Red blood cells and other nonspherical capsules in shear flow:
  {O}scillatory dynamics and the tank-treading-to-tumbling transition.
\newblock \emph{Phys. Rev. Lett.} 98:078301.

\bibitem[Succi(2001)]{Succi_LBM_2001}
Succi, S., 2001.
\newblock The {L}attice {B}oltzmann equation for fluid dynamics and beyond.
\newblock Oxford University Press, Oxford.

\bibitem[Malevanets and Kapral(1999)]{Malevanets_MSM_1999}
Malevanets, A., and R.~Kapral, 1999.
\newblock Mesoscopic model for solvent dynamics.
\newblock \emph{J. Chem. Phys.} 110:8605--8613.

\bibitem[Hoogerbrugge and Koelman(1992)]{Hoogerbrugge_SMH_1992}
Hoogerbrugge, P.~J., and J.~M. V.~A. Koelman, 1992.
\newblock Simulating microscopic hydrodynamic phenomena with dissipative
  particle dynamics.
\newblock \emph{Europhys. Lett.} 19:155--160.

\bibitem[Dao et~al.(2006)Dao, Li, and Suresh]{Dao_MBA_2006}
Dao, M., J.~Li, and S.~Suresh, 2006.
\newblock Molecularly based analysis of deformation of spectrin network and
  human erythrocyte.
\newblock \emph{Mater. Science Engin. C} 26:1232--1244.

\bibitem[Helfrich(1973)]{Helfrich_EPB_1973}
Helfrich, W., 1973.
\newblock Elastic properties of lipid bilayers: theory and possible
  experiments.
\newblock \emph{Z. Naturforschung C} 28:693--703.

\bibitem[Lidmar et~al.(2003)Lidmar, Mirny, and Nelson]{Lidmar_VSB_2003}
Lidmar, J., L.~Mirny, and D.~R. Nelson, 2003.
\newblock Virus shapes and buckling transitions in spherical shells.
\newblock \emph{Phys. Rev. E} 68:051910.

\bibitem[gri()]{gridgen}
Gridgen.
\newblock Pointwise, Inc., http://www.pointwise.com.

\bibitem[Mirijanian and Voth(2008)]{Voth_UEP_2008}
Mirijanian, D.~T., and G.~A. Voth, 2008.
\newblock Unique elastic properties of the spectrin tetramer as revealed by
  multiscale coase-grained modeling.
\newblock \emph{Proc. Natl. Acad. Sci. USA} 105:1204--1208.

\end{thebibliography}


\section{Tables}
\label{sec:tables}

\begin{table}[!h]
  \centering
\begin{tabular}{|c|c|c|c|c|}
  \hline
  Method & $d(l)$ & degree-6 & degree-5 and degree-7 & other degrees \\
  \hline
  point charges & $[0.15,0.18]$ & $90\%-95\%$ & $5\%-10\%$ & $0\%$  \\
  \hline
  Gridgen & $[0.13,0.16]$ & $45\%-60\%$ & $37\%-47\%$ & $3\%-8\%$  \\
  \hline
  free energy relaxation & $[0.05,0.08]$ & $75\%-90\%$ & $10\%-25\%$ & $0\%$  \\
  \hline
\end{tabular}
  \caption{Mesh quality for different triangulation methods.}
\label{tab:mesh_quality}
\end{table}
\begin{table}[!h]
\addtolength{\tabcolsep}{-3pt}
  \centering
\begin{tabular}{|c|c|c|c|c|c|c|c|c|c|c|c|}
  \hline
  Model & $N_v$ & $l_0^D$ & $\mu_0^D$ & $D_0^D$ & $r_c$ & $r_{mult}$ & $k_a$ & $k_d$ & $k_v$ & $\theta_0$ & $q$ or $m$ \\
  \hline
  WLC-C & $500$ & $0.56$ & $100$ & $8.267$ & $1.0$ & $2.2$ & $5000$ & $0$ & $5000$ & $6.958^o$ & $1$ \\
  \hline
  WLC-C & $1000$ & $0.4$ & $100$ & $8.285$ & $1.0$ & $2.2$ & $5000$ & $0$ & $5000$ & $4.9^o$ & $1$ \\
  \hline
  WLC-C & $3000$ & $0.23$ & $100$ & $8.064$ & $1.0$ & $2.2$ & $5000$ & $0$ & $5000$ & $2.821^o$ & $1$ \\
  \hline
  FENE-C & $500$ & $0.56$ & $100$ & $8.265$ & $1.0$ & $2.05$ & $5000$ & $0$ & $5000$ & $6.958^o$ & $1$ \\
  \hline
  WLC-POW & $500$ & $0.56$ & $100$ & $8.25$ & $1.0$ & $2.2$ & $4900$ & $100$ & $5000$ & $6.958^o$ & $2$ \\
  \hline
\end{tabular}
  \caption{RBC parameters.}
\label{tab:wlc_c_params}
\end{table}

\section{Figure Legends}

Figure 1: An element of the equilateral triangulation (left) and two equilateral triangles placed on the surface of 
a sphere of radius $R$ (right).

Figure 2: RBC sketch before and after deformation.

Figure 3: Computational results for {\em different $N_v$} (left) and {\em spring models} (right) compared with 
the experiments in \cite{Suresh_CBS_2005} and the coarse-grained (CG) RBC model in \cite{Pivkin_ACG_2008}.

Figure 4: RBC stretching along lines with {\em different orientation angles} (left) and {\em triangulation methods} 
(right) compared with the experiments in \cite{Suresh_CBS_2005}.

Figure 5: Stress-free RBC model for different triangulation methods with $N_v=500$ (left) and number of vertices with 
the energy relaxation triangulation (right) compared with the experiments in \cite{Suresh_CBS_2005}.

Figure 6: The stretching response of the stress-free RBC model for different ratio $r_{mult}$ (left) and number 
of vertices in percents which are subject to the stretching force (right) compared with the experiments in 
\cite{Suresh_CBS_2005}.

Figure 7: A single spectrin tetramer stress-strain response \cite{Voth_UEP_2008} versus the spring force of 
the spectrin-level RBC model.


\section{Figures}


\end{document}